\documentclass[pre,twocolumn,preprintnumbers,nofootinbib,amsmath,amssymb,showkeys]{revtex4}
\usepackage{graphicx}		
\usepackage{dcolumn}			
\usepackage{bm}				
\usepackage{graphics}		
\usepackage{subfig}

\begin{document}

\title{Pavlovian Prisoner's Dilemma in one-dimensional cellular automata: analytical results, the {\it quasi}-regular phase, spatio-temporal patterns and parameter space exploration}

\author{Marcelo Alves Pereira$^{1,~\dag}$ and
Alexandre Souto Martinez$^{1,~2,~\ddag}$
}

\affiliation{$^1$Universidade de S\~{a}o Paulo\\
Faculdade de Filosofia, Ci\^{e}ncias e Letras de Ribeir\~{a}o Preto\\
Av. Bandeirantes, 3900, 14040-901 \\
Ribeir\~{a}o Preto, SP, Brazil \\
$^{\dag}$marceloapereira@usp.br \\
$^2$National Institute of Science and Technology in Complex Systems \\
$^{\ddag}$asmartinez@usp.br \\
}

\date{\today}

\begin{abstract}
The Prisoner's Dilemma (PD) game is used in several fields due to the emergence of cooperation among selfish players.
Here, we have considered a one-dimensional lattice, where each cell represents a player, that can cooperate or defect.
This one-dimensional geometry allows us to retrieve the results obtained for regular lattices and to keep track of the system spatio-temporal evolution.
Players play PD with their neighbors and update their state using the Pavlovian Evolutionary Strategy.
If the players receive a positive payoff greater than an aspiration level, they keep their states and switch them, otherwise.
We obtain analitycally the critical temptation values, we present the cluster patterns that emerge from the players local interaction and we perform an exploration of paramater space.
The numerical results are in accordance to the critical temptation analitycal results, it confirms that the Pavlovian strategy foment the cooperation among the players and avoid the defection.
The system also presented a new phase in the steady state, the {\it quasi}-regular phase, where several players switch their states during round to round, but the proportion of cooperators does not alter significantly.

\end{abstract}



\maketitle
\section{Introduction} \label{introducao}
\emph{Prisoner's Dilemma} (PD) is a game where two players confront themselves and each one can either cooperate or defect.
Players receive a payoff $R$ (reward) in the case of mutual cooperation and a payoff $P$ (punishment) if they are both defectors.
If one player cooperates and the other defects, they receive $S$ (sucker) and $T$ (temptation), respectively.
These payoff values must satisfy the inequalities
$T > R > P > S$ and $T + S < 2R$ to create a dilemma~\cite{axelrod_1984}.
In a single round game the best choice is the defection, since it assures a higher payoff than cooperation, independently of the opponent decision (Nash equilibrium).
However, a local minimum occurs under mutual defection, generating the dilemma.

When the PD is played repeatedly, it is called Iterated Prisoner Dilemma (IPD).
In the computer tournament, proposed by Axelrod~\cite{axelrod_1981,axelrod_1984} to compare different strategies playing IPD, a simple strategy, with only one time step memory, called \emph{tit-for-tat} (TFT), was by far the most stable one.
The player using TFT cooperates in the first round and subsequently copies the opponent last round action.
The dilemma and the cooperation, as a profitable behavior among selfish agents, make the PD the most prominent game in the Game Theory.
It is used to model problems in several research fields~\cite{anteneodo_2002,stauffer_2004,bouchaud_2002,turner_1999}.

Here, we consider the IPD, but now each player is a cell of a one-dimesional automaton and can play with $z$ neighbors.
This geometry is equivalent to the player in regular lattices with $z$ neighbors.
In a non stochastic IPD game, during time evolution, players interact according to deterministic rules.
All players play against their respective neighbors and update their states.
This process is called \emph{round} and it is the system time unit.
After long enough, the system may reach a steady state, where the asymptotic cooperators proportion, $\rho_{\infty}$, becomes time independent.
The player state update process varies according to the adopted evolutionary strategy~\cite{pincus_1970}, namely: \emph{Darwinian Evolutionary Strategy (DES)}~\cite{nowak_1992} or \emph{Pavlovian Evolutionary Strategy (PES)}~\cite{fort_2005}, that are considered here.

In the DES, the update process uses the strategy of copying the best adapted player behavior (fittest player), also known as the ``survival of the fittest".
This is equivalent to the natural selection principle of Darwin~\cite{beyer_2002}.
The fittest player is the one who receives the greatest payoff.
Each player compares his/her own payoff to the neighboring ones, and then copy the state of the fittest neighbor.

For the PES, let us consider the following learning techniques.
Win-stay, lose-shift (WSLS) is a general learning method used for iterated decision problems of all kinds.
It was proposed by Thorndike (1911)~\cite{thorndike_1911}, assuming that actions, which yield satisfaction, will be reinforced and actions, which yield discomfort, will be weakened.
This strategy is also called \emph{Pavlov}.
Kraines and Kraines~\cite{kraines_1989} use positive and negative reinforcement to teach an individual to respond.
In the PD context, an individual is a player.
For example, in the first round, a player chooses randomly the action $C$ or $D$ (to cooperate or to defect).
He/she plays the game and evaluates the outcome.
If he/she receives a reward due to action $C$, this individuals will be more prone to keep the action $C$.
Otherwise, if he/she is punished due to the action $C$, then it will be more probable that the individual changes his/her action to $D$.
This process can be thought as the strategy {\textit ``never change a winning team''}.
If it is desirable that an individual acts like $C$, then he/she must be rewarded, or punished, repeatedly according to the player choice to reinforce the action $C$.
In the PD, under these propositons, a player keeps a given action when he/she receives a payoff $R$ or $T$ and switches if his/her payoff is $S$ or $P$.
Namely, a player keeps his/her action when playing against a cooperator and switches it when he/she confronts a defector.

Another possible way to use the Pavlov principle is to set an aspiration level (AL) to the IPD player~\cite{posch_1999}.
The payoff can be lower, equal or greater than the AL.
If they receive a payoff higher than AL, they do not change their states and switch them, otherwise.
In the PES, in general, all players have the same aspiration level.

Pavlov based strategy is very robust in situations such as: presence of noise, i.e. a player can switch his/her state at any moment, with probability $p > 0$, regardless the adopted strategy by this player (mutation)~\cite{kraines_1993};
playing against deceiving or profiteers strategies~\cite{kraines_1995};
competition for surviving, in coevolutionary games~\cite{lorberbaum_2002,nowak_1993}.
Its important features are:
it does not forgive a defection;
it exploits altruistic strategies while it is not punished with a defection;
it can correct occasional mistakes (noisy environment), this does not happen with the {\it tit-for-tat} strategy, for instance.
Nevertheless, if the Pavlovian strategy is used as WSLS, with an aspiration level, it presents a weakness: it can be exploited by defective strategies.
It seems contradictory to be robust against profiters strategies and yet allow to be exploited by defective strategies.
It happens because a given player is concerned only with his/her own payoff and does not care about the opponent payoff.

The main variable of the PD is the \emph{temptation}.
The PD order parameter is the proportion of cooperators.
When the system evolves, it passes through a transient regime and eventually reaches a steady state, which defines the phase of the system.
If the payoff values are kept constant and only $T$ is varied, the critical temptation values appear.
Critical temptation is a temptation value that yields a total payoff to the players, which force they to switch their states, generating a phase transition.
Critical temptation values depend on the adopted strategy, on the system conectivity and on the neighborhood configuration.

In this paper, we present and solve analytically the critical temptation for the IPD in one-dimensional cellular automaton with a variable number of interacting neighbors for Pavlovian Evolutionary Strategy.
The one-dimensional geometry allows us to keep track of time evolution (history in a static bidimensional image) and the steady quantitative results obtained are similar to those of square-lattices~\cite{nowak_1992}.
In Section~\ref{modelo}, we introduce the model.
In Section~\ref{critical_temptation}, we derive analytically the critical temptation values for the PES.
In Section~\ref{resultados}, we present the {\it quasi}-regular phase, which is a new phase that emerged from our numerical results.
We also present the cluster patterns that emerge during time evolution and the exploration of parameter space (temptation to defect, $T$, and initial cooperators proportion, $\rho_0$) for some connectivity values, $z$.
Final remarks are presented in Section~\ref{conclusao}.
The pattern formation given rise to the {\it quasi}-regular phase are presented in more details in the Appendix~\ref{app_pattern}.

\section{The one-dimensional model} \label{modelo}
\begin{figure}[htbp]
\centering
	\includegraphics[width=0.9\linewidth]{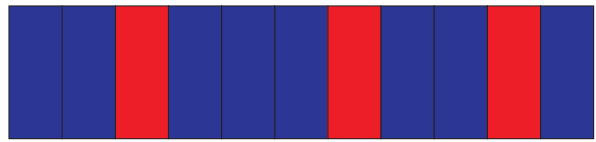}
\caption{\small
Cellular automaton in the one-dimensional lattice with $L = 11$ players and open boundary condition.
Blue cell (dark gray): cooperator, red one (light gray): defector.}
\label{fig_automato}
\end{figure}

Consider a one-dimensional cellular automaton with $L$ \emph{cells}, where each cell represents a \emph{player}.
Each player has two possible states: $\theta = 0$ (defector) or $\theta = 1$ (cooperator) (see Fig.~\ref{fig_automato}).
The automaton has no empty cells, so that $\rho_c(t) + \rho_d(t) = 1$, with
$\rho_c(t) = (1/L) \sum_{i = 1}^{L} \theta_i(t)$,
where $\rho_c(t)$ is the cooperators proportion at time $t$, and $\rho_d(t)$ is the defectors proportion.
The initial cooperators proportion, $0 \leq \rho_c(0) \equiv \rho_0 \leq 1$, is one of the problem parameters.
The position of every $L\rho_0$ cooperators in the automaton is set randomly from a uniform deviate.
The initial configuration is the only stochasticity in the model.

\begin{figure}[htbp]
\centering
\subfloat[]{
\includegraphics[width=0.9\linewidth]{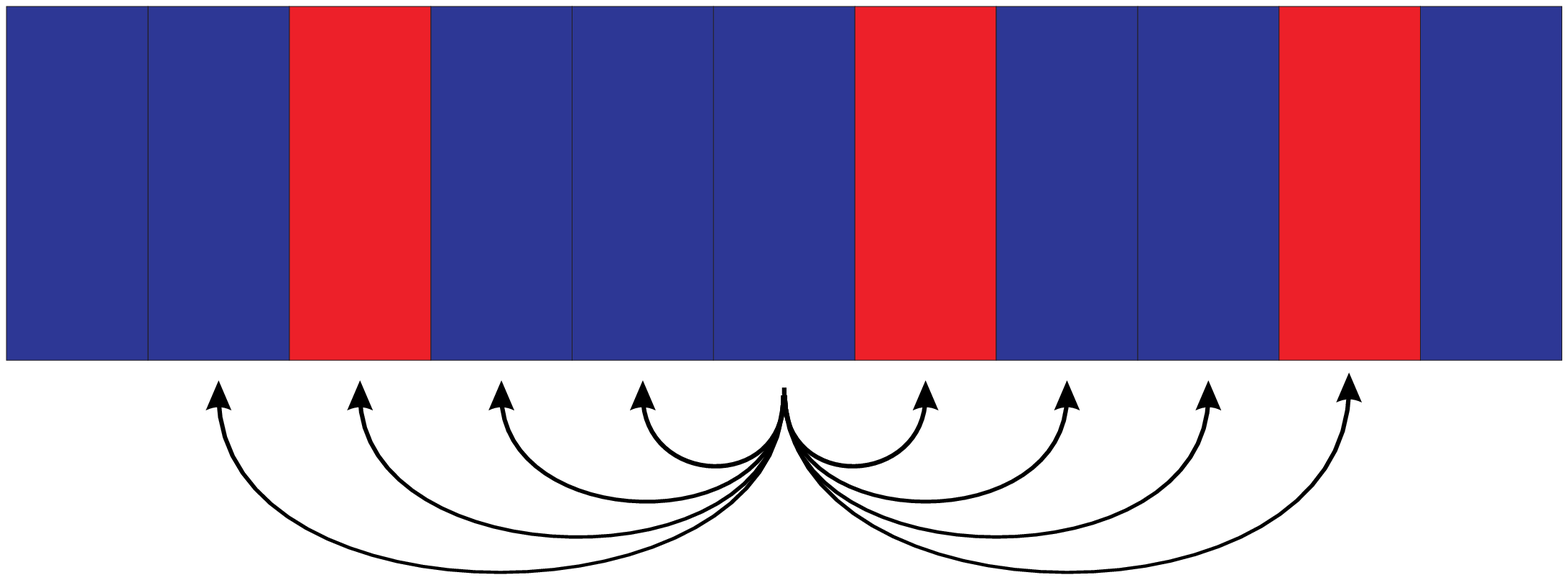}
\label{fig_conectividade-a}
} \\
\subfloat[]{
\label{fig_conectividade-b}
\includegraphics[width=0.9\linewidth]{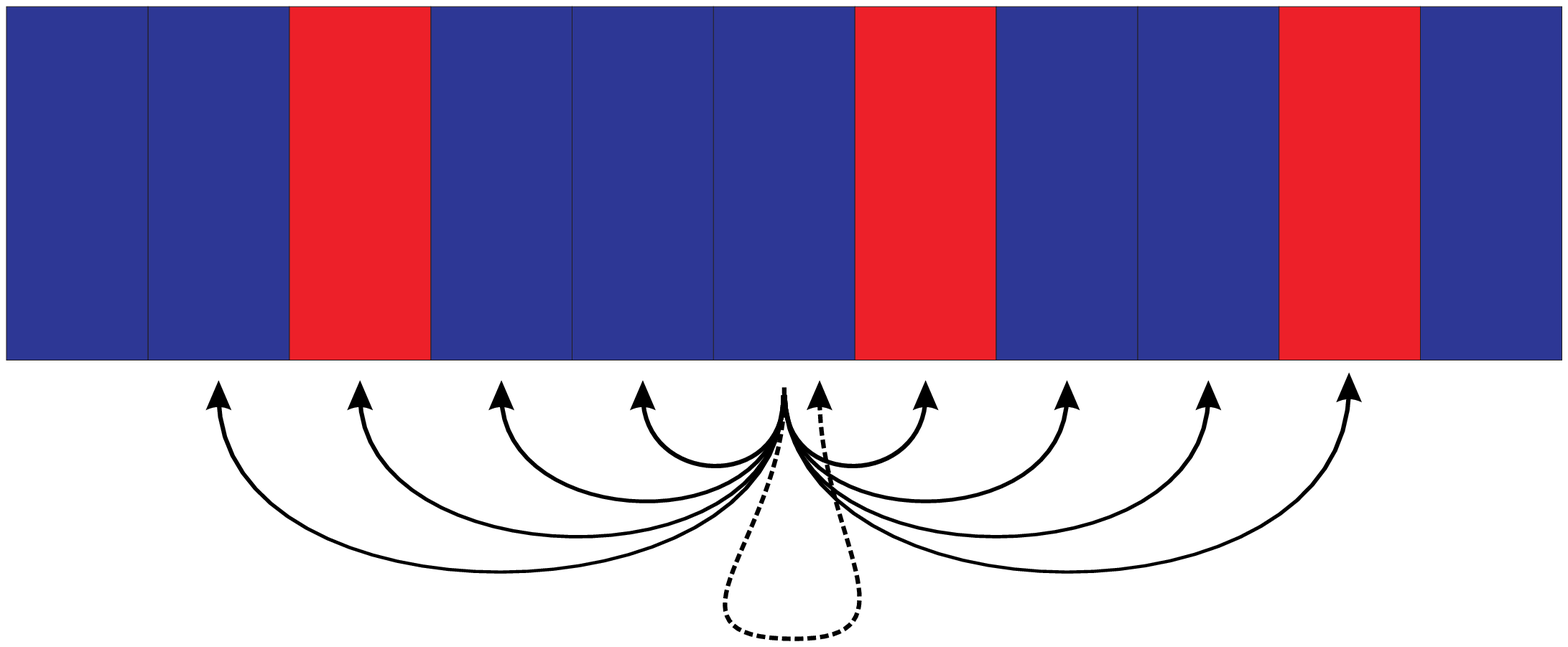}
}
\caption{\small
Cellular automaton in the one-dimensional lattice with $L = 11$ players.
The central player (arrow origin) interacts with neighbors indicated by arrows.
(a) $z = 8$ (without self-interaction).
(b) $z = 9$ (with self-interaction).
}
\label{fig_conectividade}
\end{figure}

Consider the $i$-th player, his neighborhood (or connectivity) is given by $z = \{1,~2,~\ldots,~L\}$.
If $z$ is even, there are $\alpha = z/2$ adjacent players to the right-hand side and another $\alpha = z/2$ to left-hand side (see Fig.~\ref{fig_conectividade-a}).
If $z$ is odd, each side has $\alpha = (z-1)/2$ interactive players and player $i$ interacts with his/her own state (\emph{self-interaction}) (see Fig.~\ref{fig_conectividade-b})
\cite{soares_2006,pereira_2008_IJMPC,pereira_2008_BJP}.
Nowak and May \cite{nowak_1993} argue that self-interaction makes sense, for example, if several animals (a family) or molecules can occupy a single cell.
The \emph{self-interaction} is considered an \emph{intra}-group interaction.

\begin{figure}[htbp]
\centering
\subfloat[]{
\includegraphics[width=0.45\linewidth]{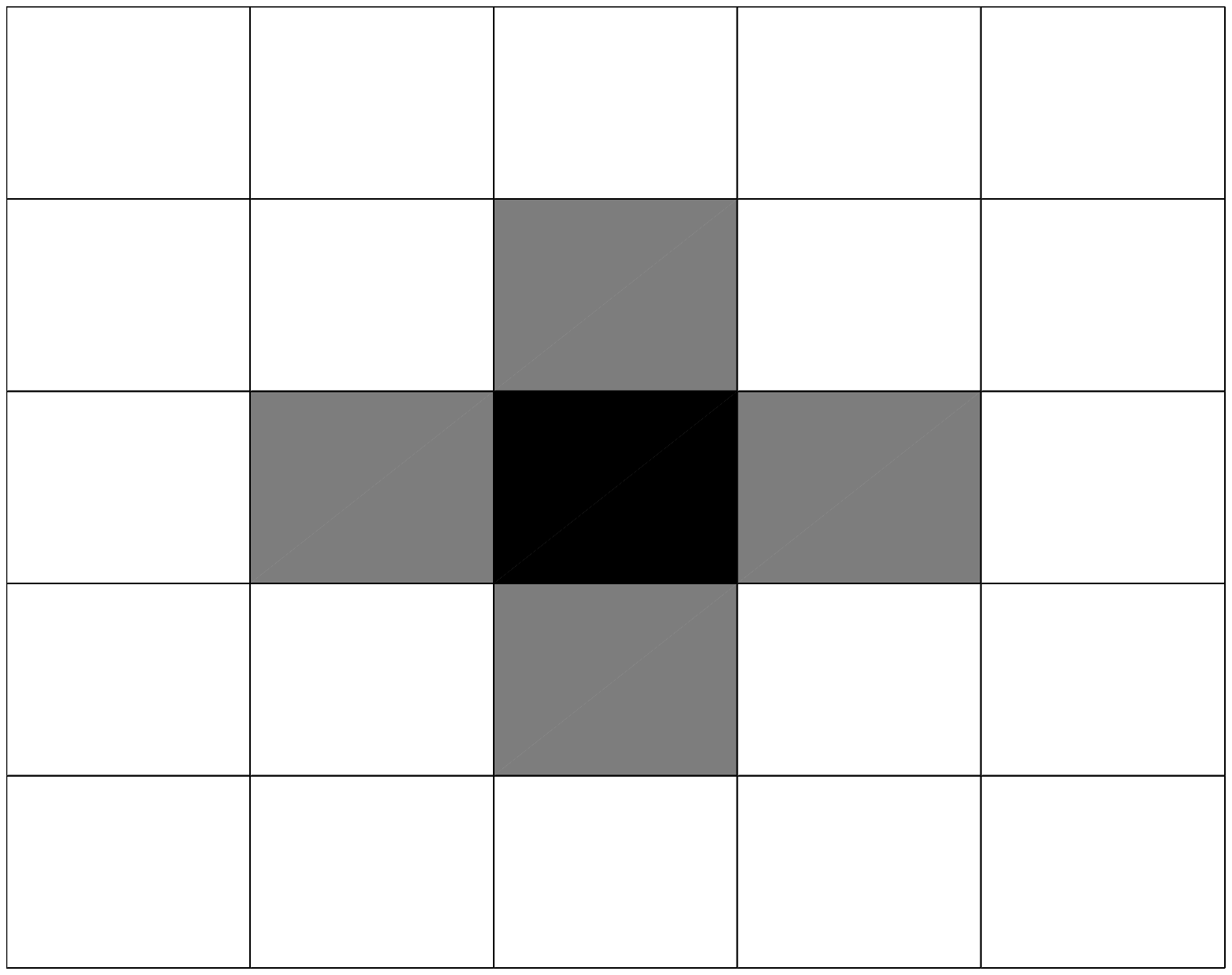}
\label{fig_vizinhança-a}
}
\subfloat[]{
\includegraphics[width=0.45\linewidth]{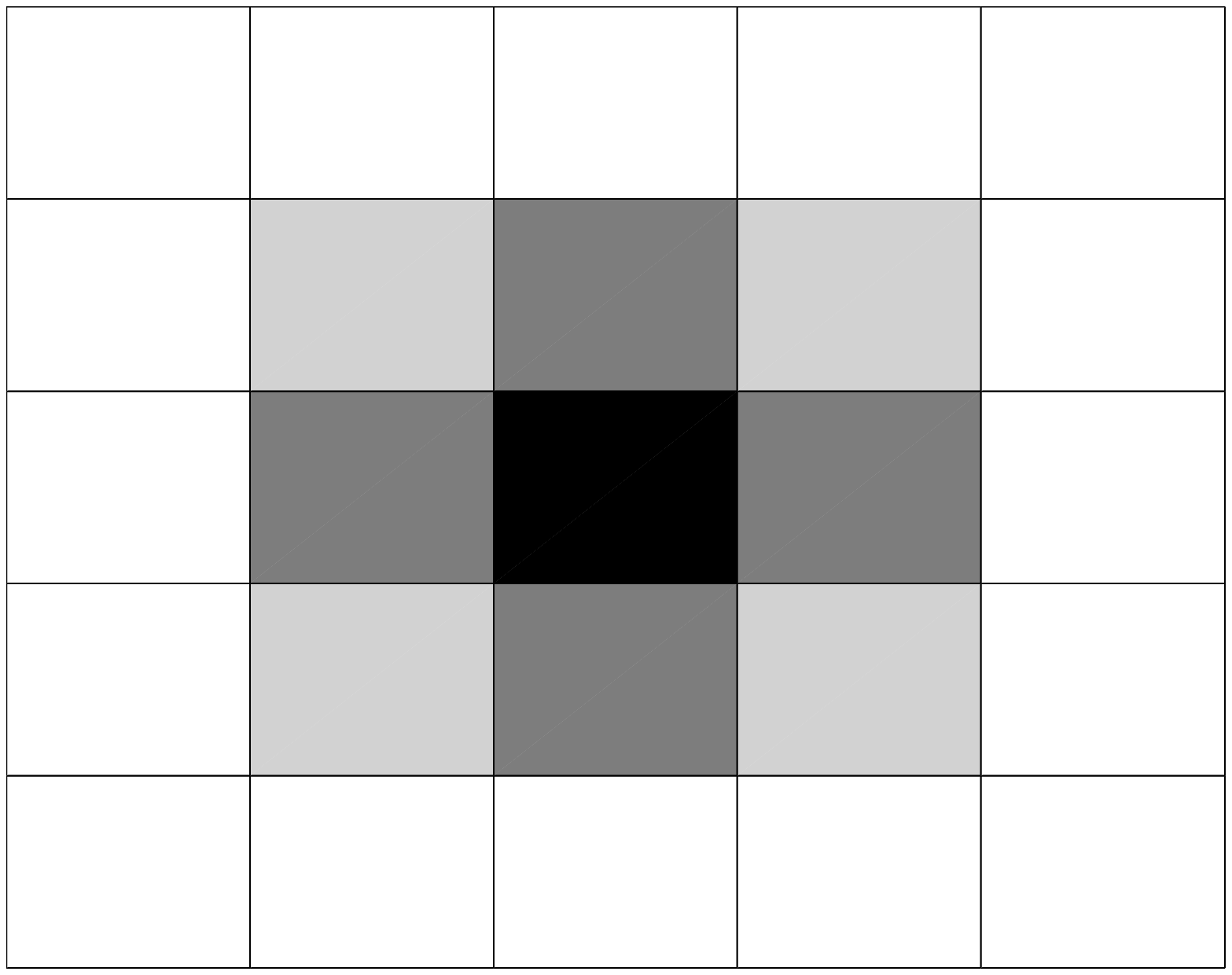}
\label{fig_vizinhança-b}
}\\
\subfloat[]{
\includegraphics[width=0.45\linewidth]{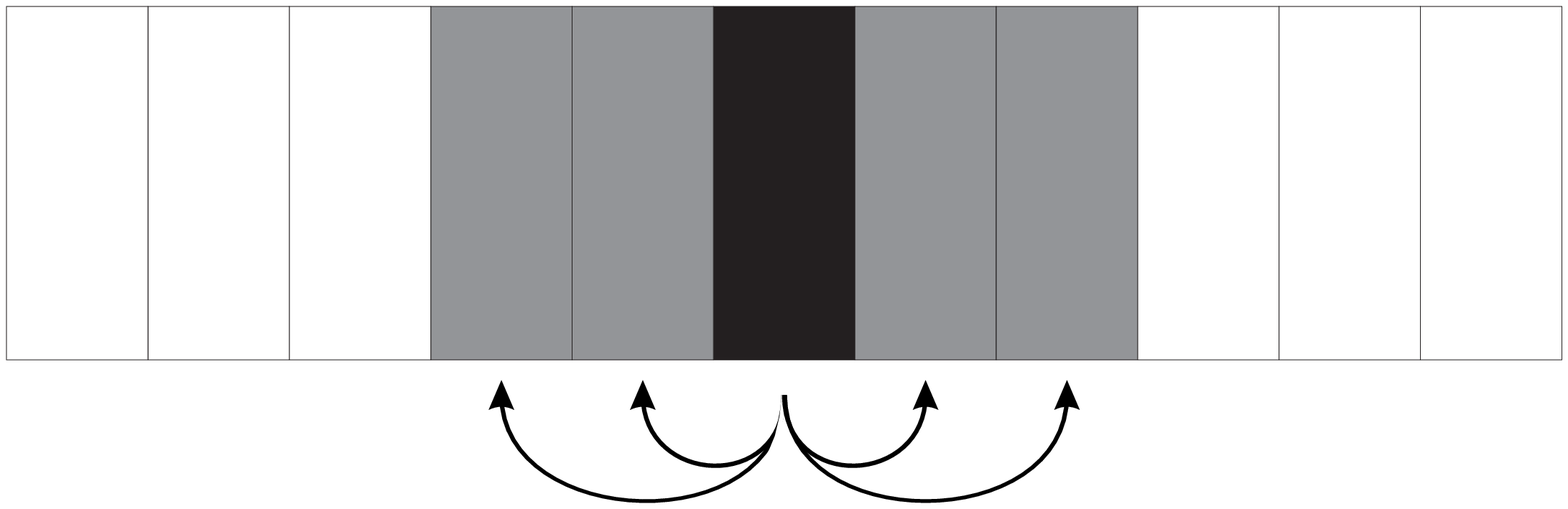}
\label{fig_vizinhança-c}
}
\subfloat[]{
\includegraphics[width=0.45\linewidth]{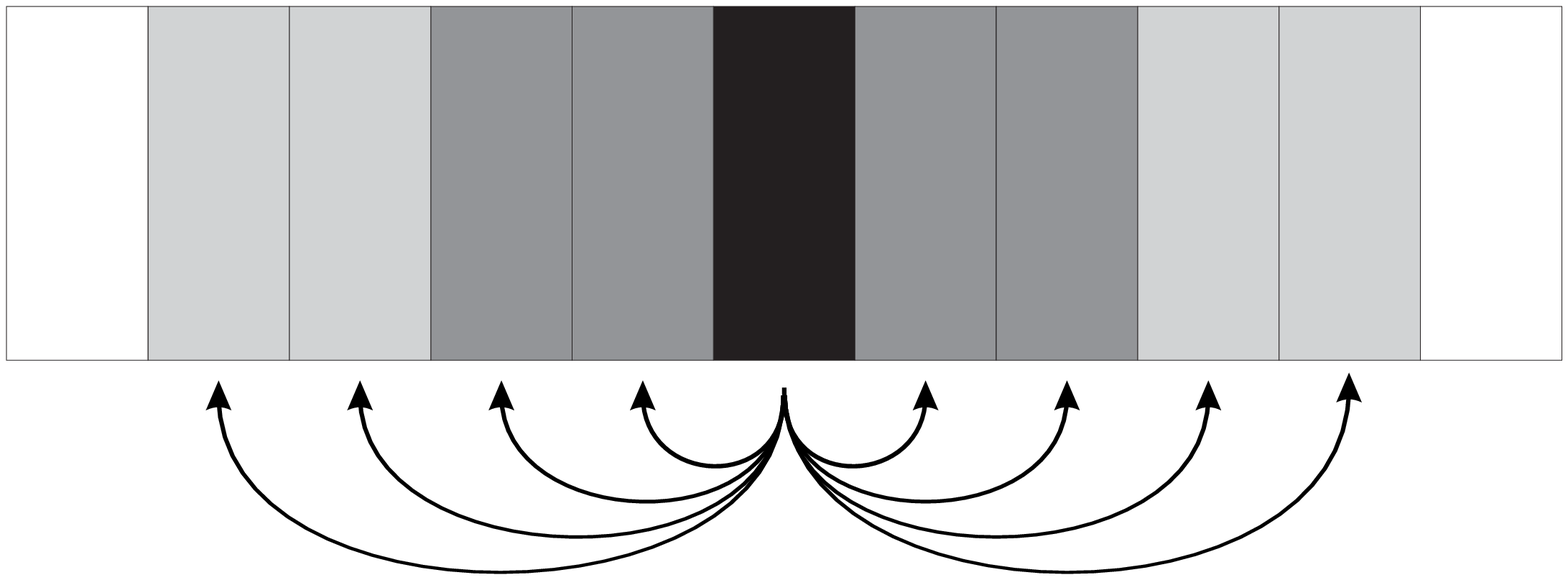}
\label{fig_vizinhança-d}
}
\caption{\small
Neighborhood representation in square lattice:
(a) von Neumman, $z = 4$ and
(b) Moore, $z = 8$.
And their representations in the one-dimensional lattice to:
(c) $z = 4$ and
(d) $z = 8$.
Black: central player; dark gray: first neighbors; and light gray: seconds neighbors.
Remember that for even $z$ there is no self-interaction.
}
\label{fig_vizinhança}
\end{figure}

Using the one-dimensional topology, it is possible to vary the lattice connectivity $z$ to any integer value in the range $1 \leq z \leq L$.
This is not possible, for instance, in a square lattice, because it is limited to von Neumman ($z = 4$, see Fig.~\ref{fig_vizinhança-a}) or Moore ($z = 8$, see Fig.~\ref{fig_vizinhança-b}) neighborhoods.
In a square lattice, if $z$ is different from $z = \{4;~8;~24\}$, the neighborhood is asymmetric.
For instance, to obtain $z = 6$, one must consider the honeycomb lattice.
Since the critical temptation values depend only on the coordination number, this neighborhood may be considered in a one-dimensional lattice, where $z = \{4;~5\}$ corresponds to the von Neumman neighborhood, $z = \{8;~9\}$ matches the Moore one and $z = \{6;~7\}$ the honeycomb case, with and without self-interaction, respectively.
We have used periodic boundary conditions (PBC), every player has the same connectivity.
Once the lattice is one-dimensional, the boundary effect is smaller than observed in $d$ dimensional lattices~\cite{pereira_2008_IJMPC,pereira_2008_BJP}.

\section{Analytical calculation of Critical Temptation} \label{critical_temptation}
To present the Pavlov critical temptation calculation, we first briefly review the results obtained by Dur\'{a}n and Mulet~\cite{duran_2005} for the Darwinian strategy.
For the payoff evaluation, consider the parameters: $T$, $R$, $P$ and $S$.
Consider two players $i$ and $j$ playing PD in a cellular automaton.
The player $i$ payoff with respect to player $j$ is:
\begin{eqnarray}
	g_{i,j} & = & T[(1-\theta_i)\theta_j] + R[\theta_i \theta_j] + \nonumber \\
	& & P[(1-\theta_i)(1-\theta_j)] + S[(1-\theta_j)\theta_i],
\label{eq_ganho_generalizado}
\end{eqnarray}
where $\theta_k$, is the player $k$ state, with $k = \{1;~2; \ldots;~L\}$.
The total $i$-th player payoff is:
$G_i = \sum_{j = 1}^{z} g_{i,j}$.
It is noteworthy that, only if $z$ is odd, there is an extra payoff component $g_{i,i}$, due to self-interaction.
From Eq.~\ref{eq_ganho_generalizado}, the player $i$ payoffs due to the interaction with a single defector ($\theta_j = 0$) and a single cooperator ($\theta_j = 1$), are:
\begin{equation}
	g_{i,j} = 
\begin{cases}
	P(1-\theta_i) + S\theta_i & \text{if $\theta_j = 0$,} \\
	T(1-\theta_i) + R\theta_i & \text{if $\theta_j = 1$.}
\end{cases}
\end{equation}
The player $i$ payoff, due to interactions with $c_i$ cooperators within the $z$ neighbors, is:
$G_{i}^{(c_i)}(\theta_i) = [T(1-\theta_i) + R\theta_i]c_i$, and with $d_i$ defectors:
$G_{i}^{(d_i)}(\theta_i) = [P(1-\theta_i) + S\theta_i]d_i$.
The payoffs sum, due the interactions with all the $z$ neighbors, leads to the $i$-th player total payoff:
$G_{i}(\theta_i) = [T(1-\theta_i) + R\theta_i]c_i + [P(1-\theta_i) + S\theta_i]d_i$.
Since the number of cooperators and defectors in a given neighborhood are complementary, $d = z - c$:
\begin{eqnarray}
	G_{i}(\theta_i) & = & Tc_i + P (z - c_i) + \nonumber \\
	& & [(R-T)c_i + (S - P)(z-c_i)]\theta_i.
\label{eq_ganho_total_generalizado}
\end{eqnarray}
Therefore, the player $i$ total payoff, is given by:
\begin{equation}
	G_{i}(\theta_i) =
\begin{cases}
	Tc_i + P(z-c_i) & \text{if $\theta_i = 0$,} \\
	Rc_i + S(z-c_i) & \text{if $\theta_i = 1$.}
\end{cases}
\label{eq_ganho_total_simplificado}
\end{equation}

In the following, we determine the payoffs for the DES and PES and the critical temptation values, which depend on the adopted strategy.

\subsubsection{Darwinian Evolutionary Strategy (DES)}
Nowak and May \cite{nowak_1992} used the parameters set $R = 1$, $P = S = 0$, leaving only one free parameter, the temptation $1 \leq T \leq 2$, that ensures the conflict conditions.
These values are different from those originally defined by Tucker~\cite{dresher_1961} ($T = 5$, $R = 3$, $P = 1$, $S = 0$).
The conditions $T > R > P > S$ and $T + S < 2R$ have been relaxed ($P = S$; for $T = 1$, $T = R$; and for $T = 2$, $T + S = 2R$) without any harm to the DP conflict features.
This modification is known as the Weak Prisoner Dilemma.
Placing the values adopted by Nowak and May for these parameters in Eq.
\ref{eq_ganho_generalizado}, the payoff becomes: 
$g_{i,j} = T(1-\theta_i)\theta_j + \theta_i \theta_j$.
A similar result has been obtained by Dur\'{a}n and Mulet \cite{duran_2005}:
$g_{i,j} = T(1-\theta_i \theta_j)\theta_j + \theta_i \theta_j.$
The difference between our result and the Dur\'{a}n and Mulet one is the presence of $\theta_j$ multiplying $\theta_i$ inside the parentesis, which it is unnecessary and, in this case, it does not alter the result.

However, we notice that, if the players state can assume rational values, Dur\'{a}n and Mulet result is not valid.
This situation occurs in the Continuous Prisoner's Dilemma (CPD)~\cite{ifti_2004,nowak_1999b,nowak_1999c}, where a player has a cooperation level (CL), with $0 \leq CL \leq 1$, instead of only defecting or cooperating.
For the CPD our results give the correct payoff values, considering the linear interpolation for intermediate values.

Considering $R = 1$, $P = S = 0$ in Eq.~\ref{eq_ganho_total_generalizado}, we have $G_{i}(\theta_i) = [T-(T-1)\theta_i]c_i$.
Notice that:
(i) the payoff for a cooperator who plays with $c_i$ cooperators is $G_{i}^{(c_i)}(\theta_i = 1) = c_i$; while,
(ii) a defector who plays with $c_i$ cooperators has a payoff equal to $G_{i}^{(c_i)}(\theta_i = 0) = c_iT$.
For $T > 1$:
(i) $G_{i}^{(c_i)}(\theta_i =0) > G_{i}^{(c_i)}(\theta_i = 1)$; and
(ii) $G_{i}^{(c)}(\theta) \geq G_{i}^{(c - 1)}(\theta)$.
In DES, the payoff of each player is always non-negative, $G_i \geq 0$.
After all players calculate their payoffs, they update their states.
During this process, each player $i$ compares his/her payoff $G_i$ with $G_k$, where $G_k$ is the payoff of his/her $k$-th neighbor, with $k = \{1;~2;~\ldots;~z\}$.
If $G_i < G_k$ and $G_k = \max[G \in z]$, player $i$ replicates the player $k$ state, otherwise he/she maintains his/her current state.

The system evolves till it eventually reaches the steady state, where the cooperators proportion $\rho_\infty$ is stationary.
The $\rho_\infty$ phase transitions occur when the temptation value passes through critical values $T_c$.
In the conflict region, $1 <T <2$, these transitions have been calculated~\cite{duran_2005}:
$T_c(n,m) = (z-n)/(z-n-m)$, where $0 \leq n < z$ and $1 \leq m \leq \mbox{int}[(z-n-1)/2]$ are integers\footnote{For $x$ positive, the function $\mbox{int}(x)$ gives the largest integer less than or equal to $x$.}.
For example, for $z = 8$, these values are $T_c = (8/7,~8/6,~8/5,~8/4)$.

\subsubsection{Pavlovian Evolutionary Strategy (PES)}
Following the same reasoning line as used for the DES, we present for the first time the critical temptation values calculation for the PES.
The parameters used are $P = -R$ and $S = -T$, which are placed in Eq.~\ref{eq_ganho_total_simplificado}:
\begin{equation}
	G_{i}(\theta_i) =
\begin{cases}
	Tc_i - R(z-c_i) & \text{if $\theta_i = 0$,} \\
	Rc_i - T(z-c_i) & \text{if $\theta_i = 1$.}
\end{cases}
\label{eq_pavlov_ganho_total_simplificado}
\end{equation}

For system using PES, each player payoff can be either positive or negative in the range: $-zT < G_i < zT$ (Eq.~\ref{eq_pavlov_ganho_total_simplificado} extreme cases are: $c_i = 0$ and $c_i = z$).

Each player $i$ evaluates his/her payoff $G_i$.
If the payoff is greater than the aspiration level ($G_i > AL$, with $AL = 0$), the player maintains his/her current state, otherwise, he/she switches the current state.
We have defined the aspiration level as a null payoff, but any other value can be choosen.

For player $i$ to switch his/her state, it is necessary that his/her payoff be null or negative, that is:
$G_{i}(\theta_i) \leq 0$.
Applying this condition to the null payoff situation ($G_{i}(\theta_i) = 0$) in Eq.
\ref{eq_pavlov_ganho_total_simplificado}, one has:
\begin{equation}
	G_{i}(\theta_i) =
\begin{cases}
	c_iT - (z-c_i)R \leq 0 & \text{if $\theta_i = 0$,} \\
	c_iR - (z-c_i)T \leq 0 & \text{if $\theta_i = 1$.}
\end{cases}
\label{eq_pavlov_ganho_condicionado}
\end{equation}

For a defector to maintain his/her current state, $T$ must provides a null gain: $c_iT_c - (z-c_i)R = 0$, which leads to critical temptation value:
$T_c = [(z-c_i)/c_i]R$,
and in a cooperator case, the null payoff occurs when $c_iR - (z-c_i)T_c = 0$ and
$T_c = [c_i/(z-c_i)]R$.
These two cases can be written by a simple equation:
\begin{equation}
	T_c(z,c_i) = \left(\frac{z-c_i}{c_i}\right)^{(-1)^{\theta_i}}R.
\label{eq_pavlov_Tc_generalizado}
\end{equation}
The relevant variable is $[(z-c_i)/c_i]^{\theta_i}$, which strongly contrasts to the DES one: $(z-n)/(z-n-m)$.
However, notice that, as the DES, it does not depend on the configuration of the $c_i$ cooperators within the $z$ neighbors, it depends only on the the $c_i$ and $z$ values.
For this reason, we can use the one-dimensional geometry in the following.

An interesting feature observed for $T_c$ in PES is its dependence on the player state.
Critical temptation values are the same for defectors and cooperators, but they appear in reverse order.
For example, consider a cooperator playing against $z = 4$ neighbors, if in the neighborhood there is no cooperator, $T_c(4,0) = 0$, one cooperator, $T_c(4,1) = 1/3R$, and so on, then $T_c(4,c_i) = \{0;~1/3R;~1/2R;~3R;~\infty\}$, for $c_i = \{1,~2,~3,~4\}$.
Now consider a defector in the same situation, $T_c(4,c_i) = \{\infty;~3R;~1/2R;~1/3R;~0\}$, for $c_i = \{1,~2,~3,~4\}$.

\section{Numerical results: emergence of the new {\it quasi}-regular phase} \label{resultados}
We have written a numerical code to simulate the PD adopting the PES in one-dimensional cellular automaton.
The parameters are: $L=1,000$ cells, with $L\rho_0$ being the number of cooperators and the remaining ones the number of defectors.
To be statistically meaningful, the asymptotic proportion of cooperators, $\rho_\infty$, are averages obtained for $1,000$ realizations.
The quantity $T$ varies in steps $\Delta T = 0.01$ in the range $1 \leq T \leq 2$ and $\rho_0$ varies in steps $\Delta \rho_0 = 0.1$ in the range $0 < \rho_0 < 1$.
The spatio-temporal patterns yielded by the cooperative/defective clusters are presented using smaller systems than the used to calculate $\rho_\infty$.

Despite the equivalence with $d$ dimensional lattices in the critical temptation values determination, the one-dimensional case has several advantages~\cite{pereira_2008_IJMPC,pereira_2008_BJP}: it is easier to explain the cooperative/defective clusters invasion process and also the $\rho_\infty$ oscillations during the steady regime as observed in the Nowak and May pioneer work~\cite{nowak_1992, nowak_1993}.
In addition to these phenomena explanation, it is also possible to save the system history in a single static image (see the spatio-temporal patterns).
This is impossible to perform in a two-dimensional system, for example, where it is necessary a movie to observe the time evolution.

In  the steady state, the system can reach the cooperative, chaotic or defective phases, when adopting the DES and the cooperative or {\it quasi}-regular phases (which was not characterized before), adopting the PES.
The cooperative phase is characterized by the majority of players being cooperators.
If the majority of players are defectors, the system is in the defective phase.
These two phases are not sensible to the initial configuration.
In these cases the fluctuations of $\rho_\infty$ (standard deviation - SD) is almost null.
In contrast, the chaotic phase is highly sensitivity to small changes in the initial configuration (larger $\rho_\infty$ fluctuations - $SD \sim 0.5$).

In the {\it quasi}-regular phase, $\rho_\infty$ oscillates a little around $\rho_\infty \sim 0.5$, however there is a very large number of players, who switch their states.
However, these switching balance themselves.
And the system is not sensible to the initial configuration presenting a small $SD$, with $SD \sim 0$ over almost all the parameter space.

After the transient regime, the system reaches the steady state with the asymptotic $\rho_\infty$.
Adopting the PES, the system can present only the cooperative or {\it quasi}-regular phases.
The defective and chaotic phases are absent with the PES.
The defective phase does not occur because a defective cluster yields negative payoff to its members.
Thus, they change their states when this happens.
The chaotic phase absence is confirmed by the small standard deviation, $SD \sim 0$ all over the parameter space.

In the following we show the phase-diagram of systems adopting the PES, where are present the cooperative and {\it quasi}-regular phases and the patterns that emerge during the transient time and the ones which persist in the steady state.
The patterns are a visual way to understand the phases.

\subsection{Transient and Steady Regimes: exploration of the parameter space}
\label{resultados_pavlov_estacionario}
\begin{figure}[htbp]
\subfloat[$\rho_\infty$ for $z = 2$.]{
\label{fig_pav_superficie_02-a}
\includegraphics[width=0.45\linewidth]{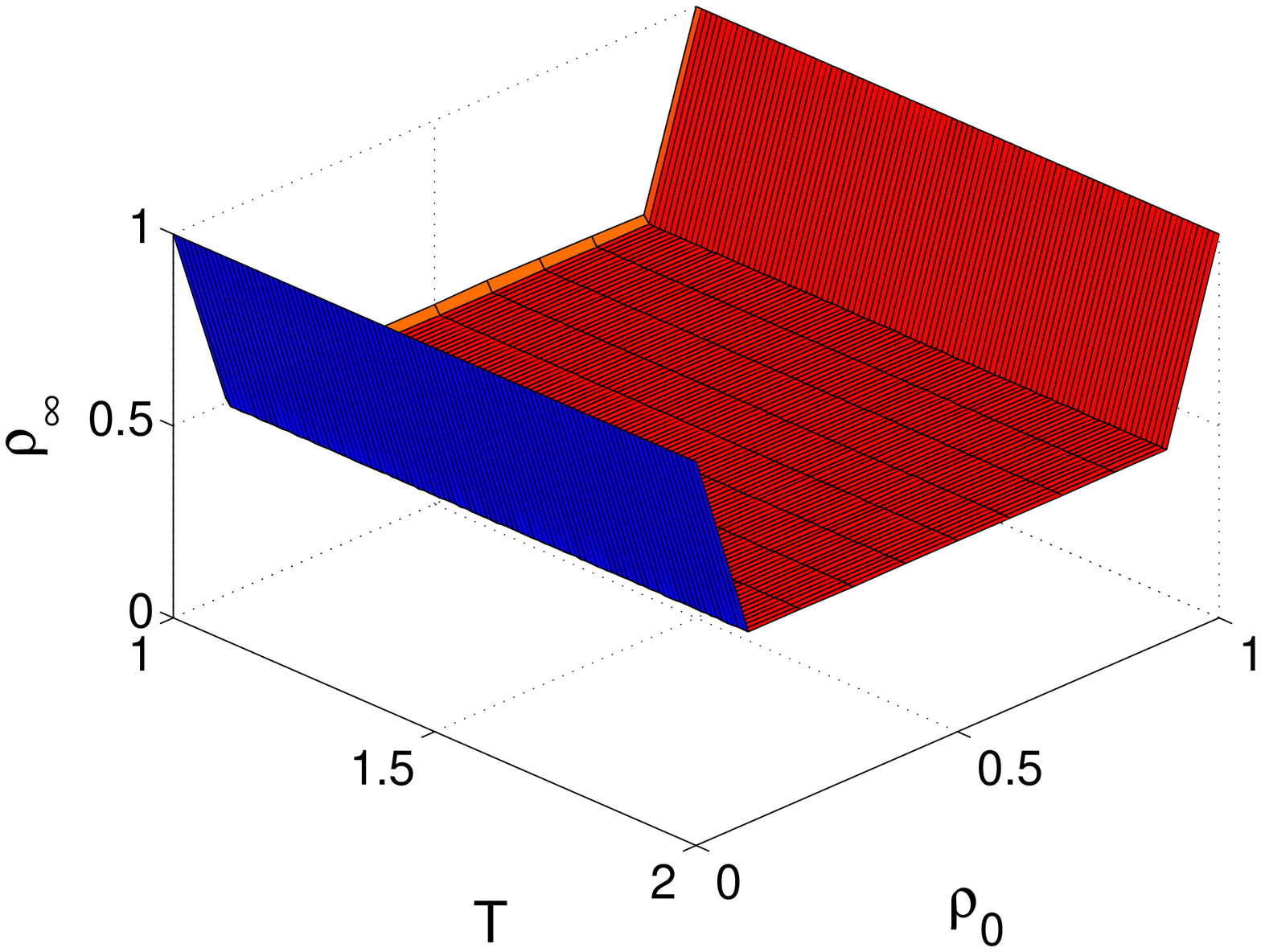}
}
\subfloat[$\rho_\infty$ standard deviation for $z = 2$.]{
\label{fig_pav_superficie_02-b}
\includegraphics[width=0.45\linewidth]{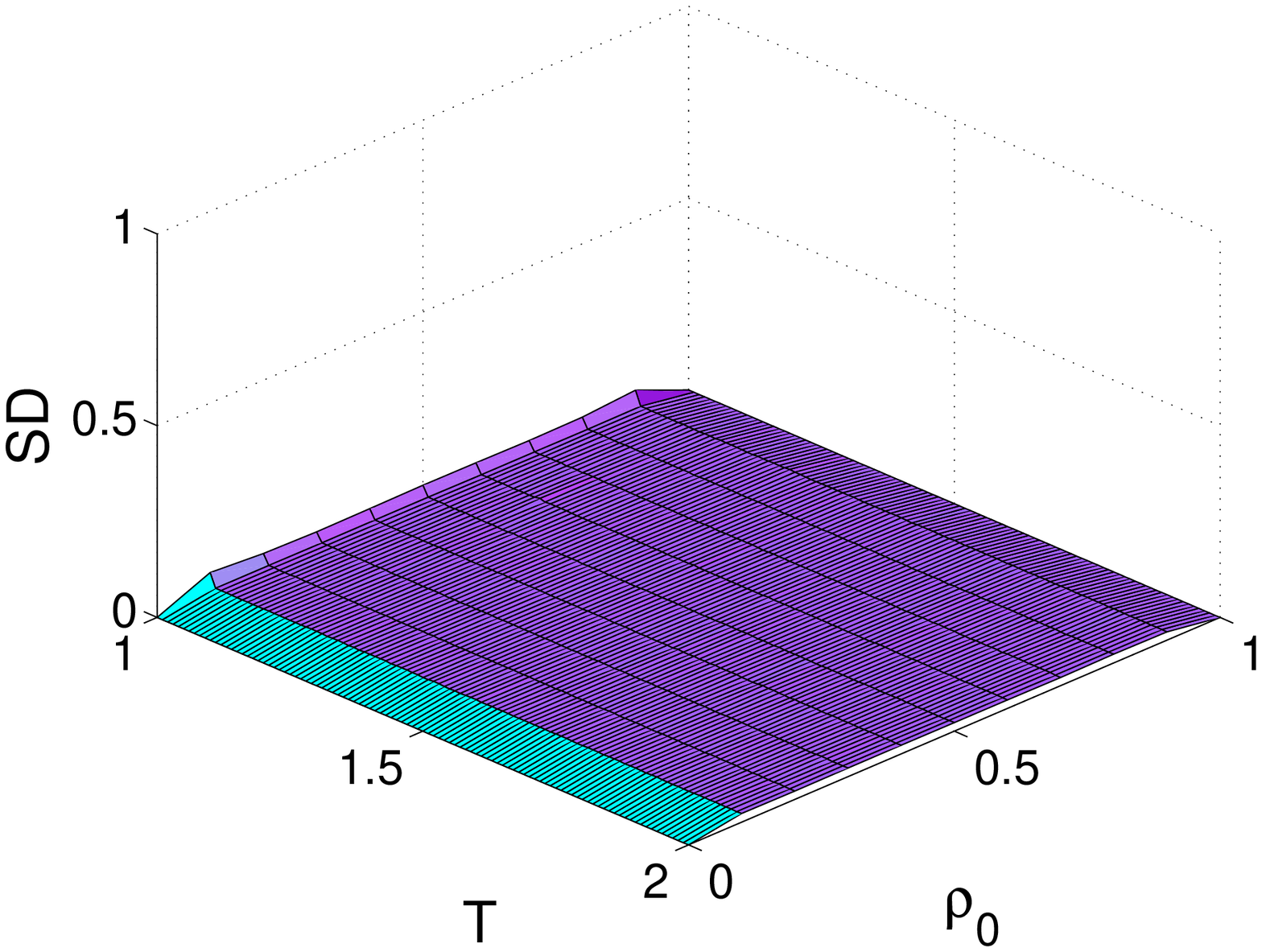}
}
\\
\subfloat[$\rho_\infty$ for $z = 8$]{
\label{fig_pav_superficie_08-a}
\includegraphics[width=0.45\linewidth]{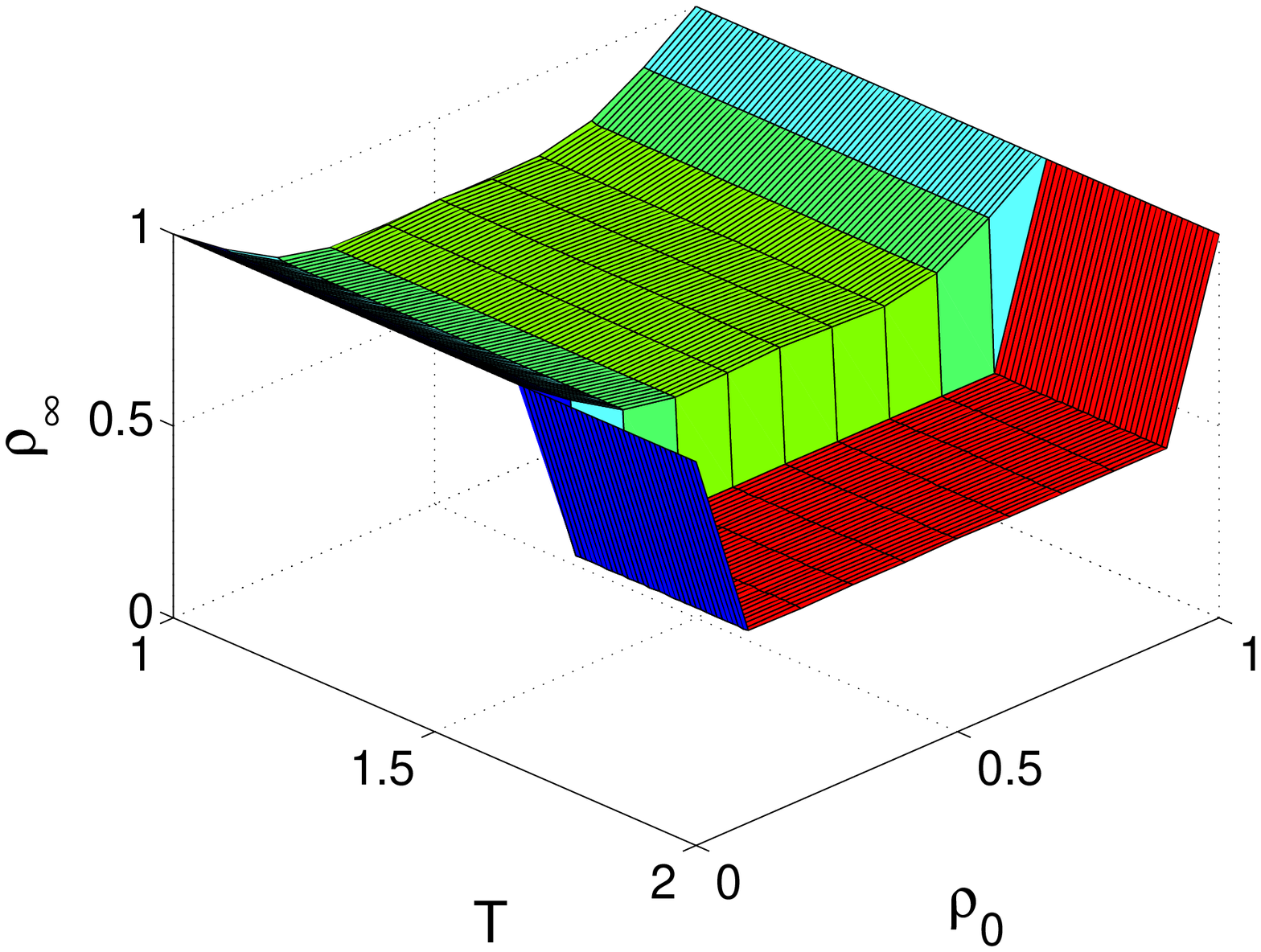}
}
\subfloat[$\rho_\infty$ standard deviation for $z = 8$.]{
\label{fig_pav_superficie_08-b}
\includegraphics[width=0.45\linewidth]{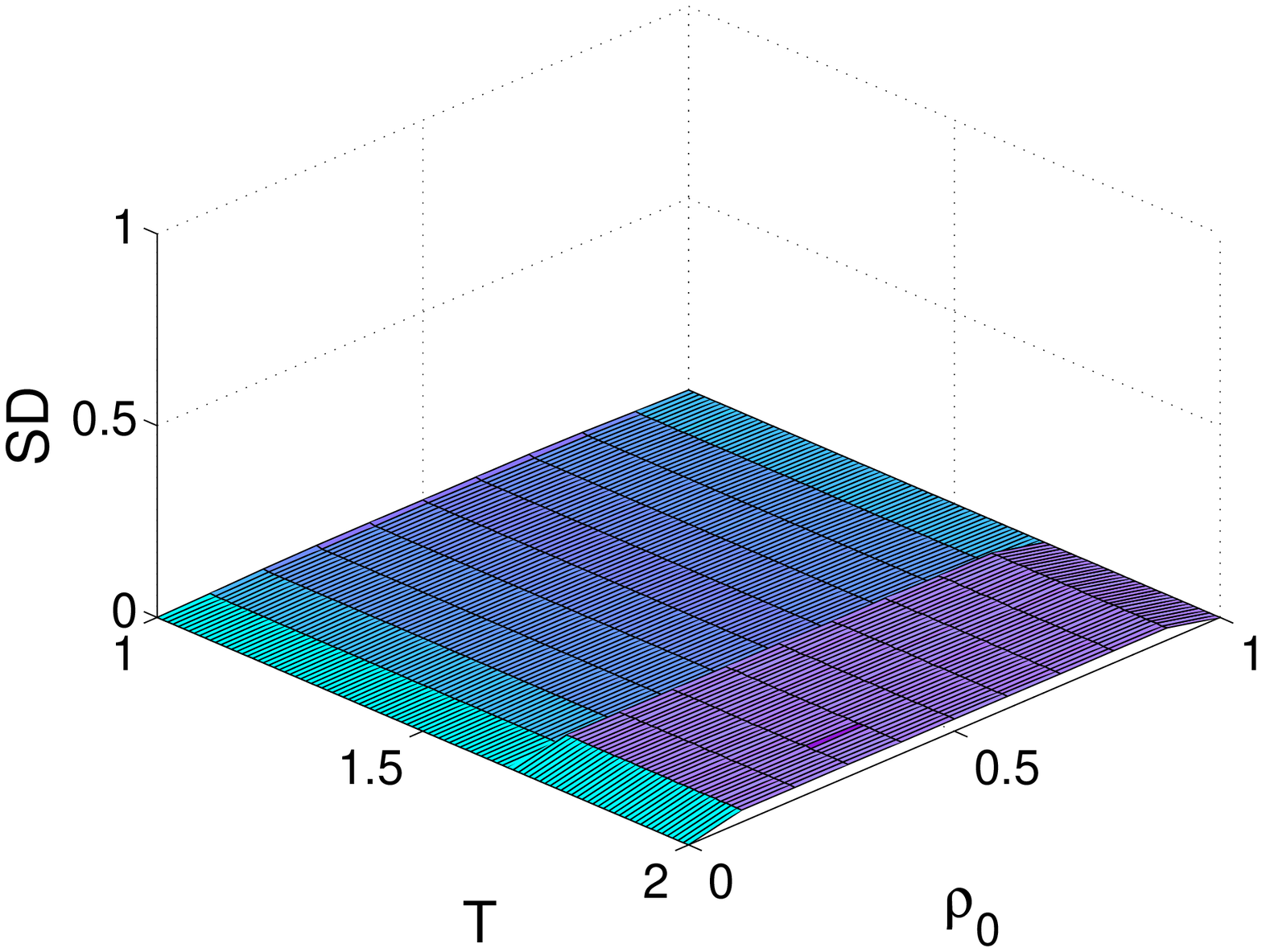}
}
\\
\subfloat[$\rho_\infty$ for $z = 30$]{
\label{fig_pav_superficie_30-a}
\includegraphics[width=0.45\linewidth]{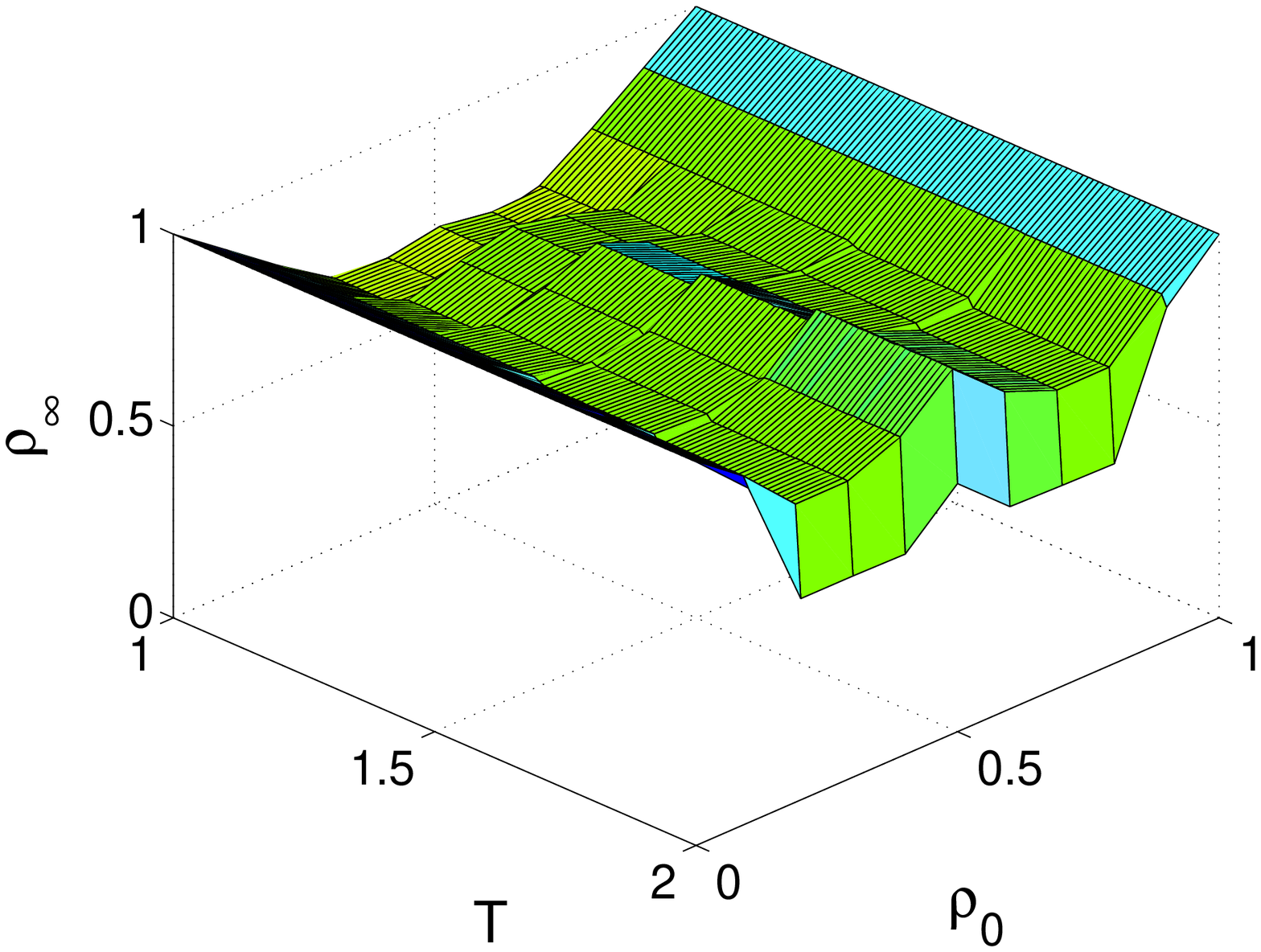}
}
\subfloat[$\rho_\infty$ standard deviation for $z = 30$.]{
\label{fig_pav_superficie_30-b}
\includegraphics[width=0.45\linewidth]{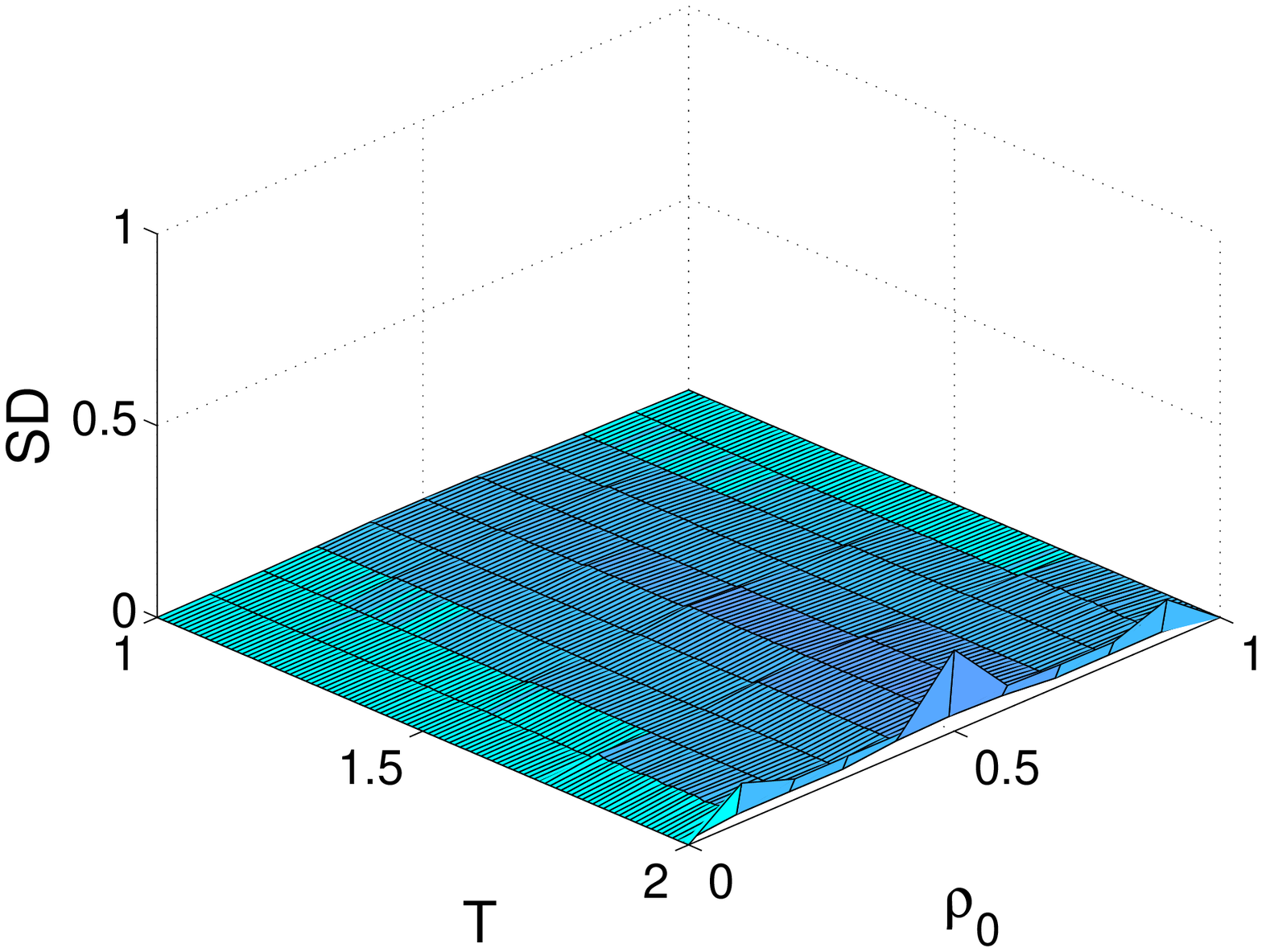}
}
\caption{
$\rho_\infty$ surface as function of the temptation value, $T$, and the initial cooperators proportion, $\rho_0$, for $z = \{2;8;30\}$.
}
\label{fig_pav_superficie_par}
\end{figure}

\begin{figure}[htbp]
\subfloat[$\rho_\infty$ for $z = 3$.]{
\label{fig_pav_superficie_03-a}
\includegraphics[width=0.45\linewidth]{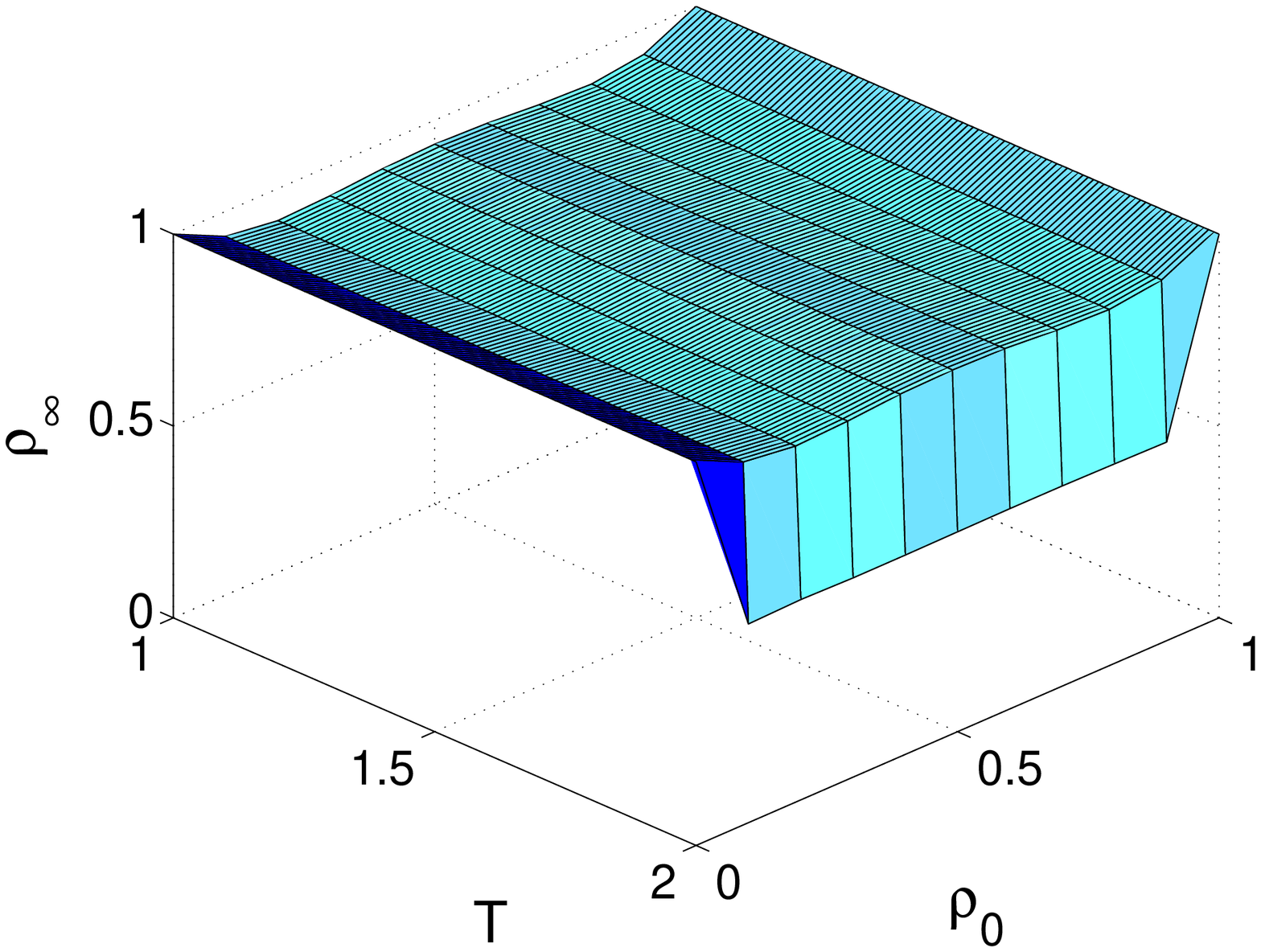}
}
\subfloat[$\rho_\infty$ standard deviation for $z = 3$.]{
\label{fig_pav_superficie_03-b}
\includegraphics[width=0.45\linewidth]{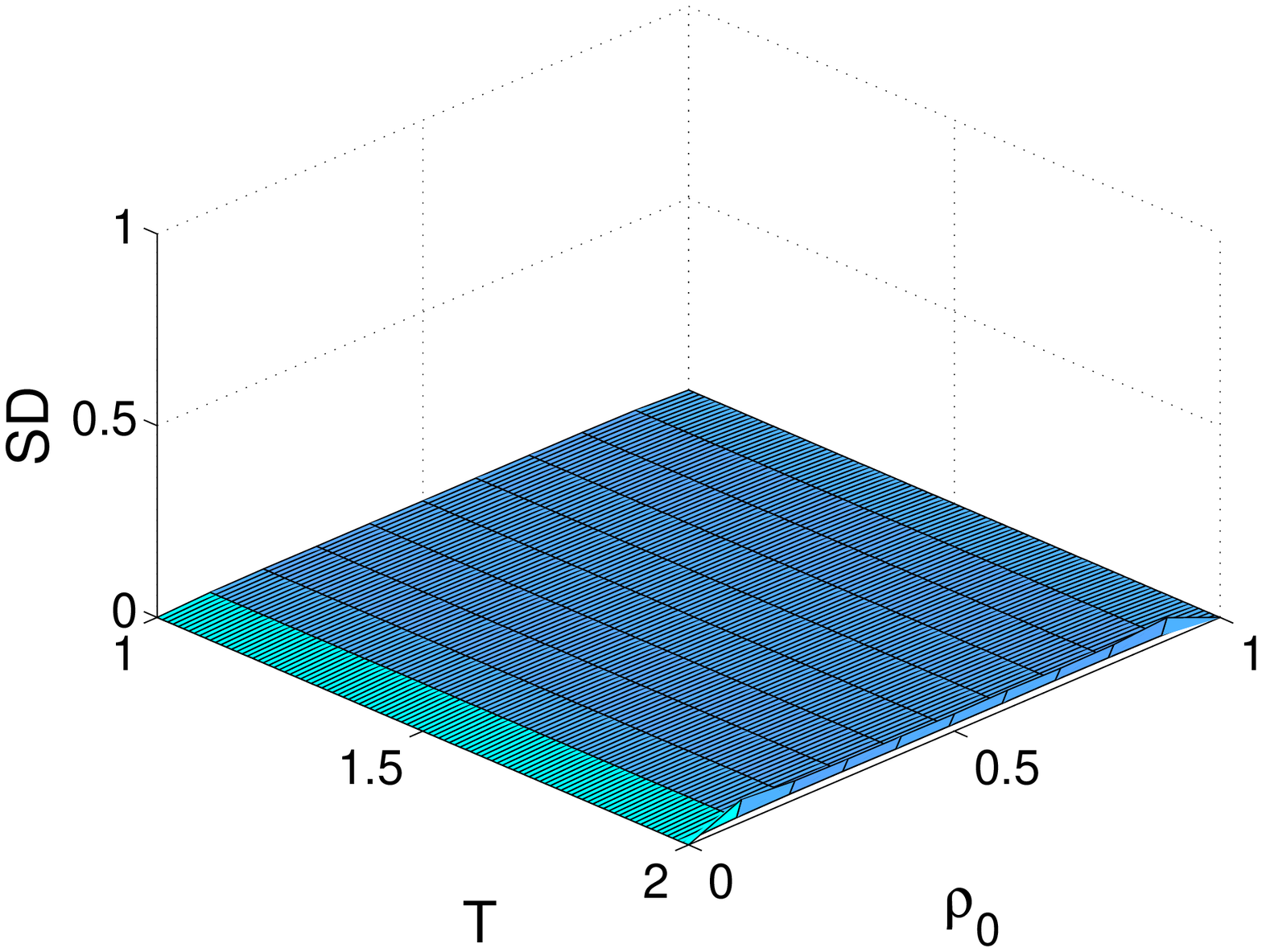}
}
\\
\subfloat[$\rho_\infty$ for $z = 9$]{
\label{fig_pav_superficie_09-a}
\includegraphics[width=0.45\linewidth]{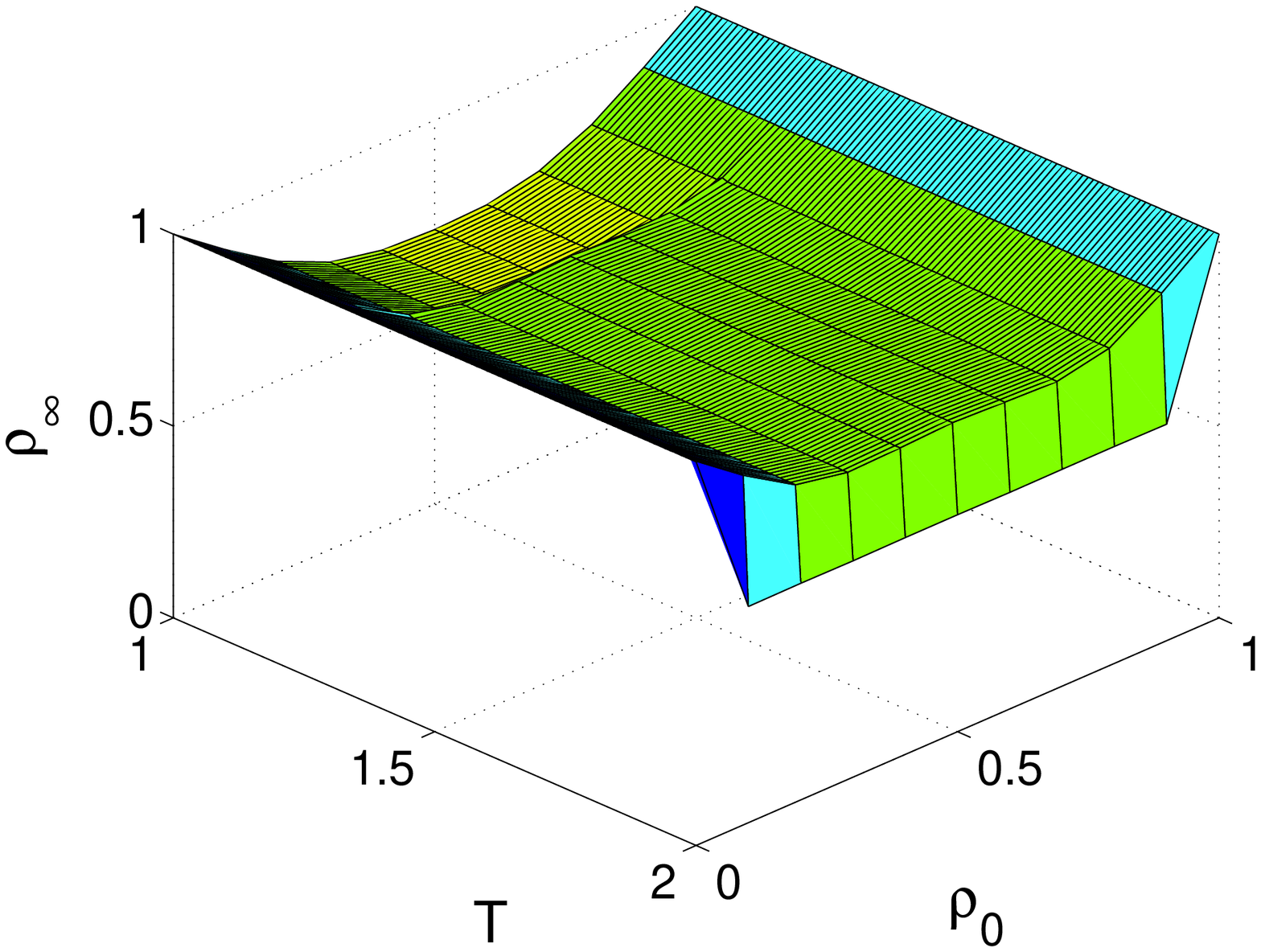}
}
\subfloat[$\rho_\infty$ standard deviation for $z = 9$.]{
\label{fig_pav_superficie_09-b}
\includegraphics[width=0.45\linewidth]{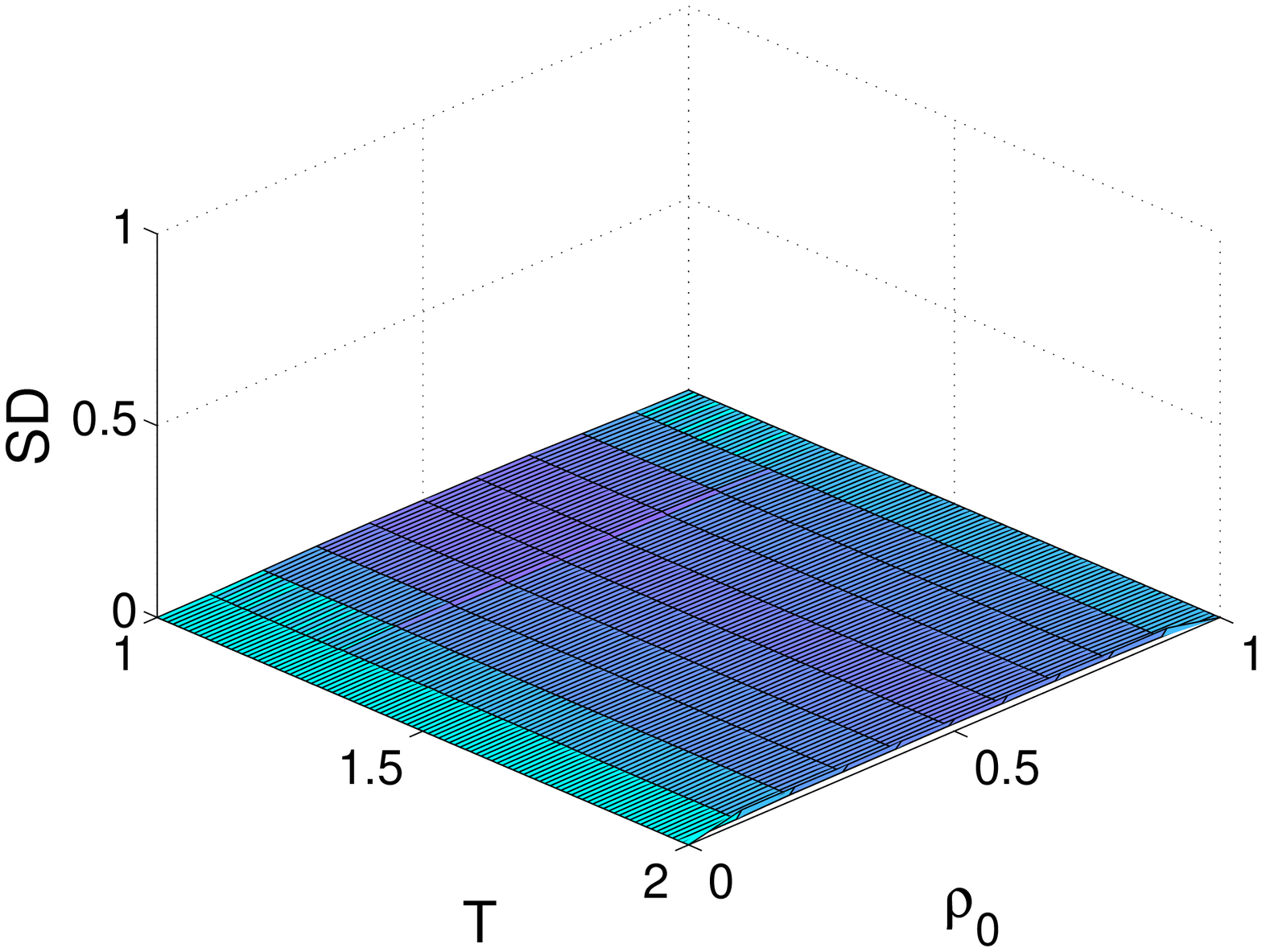}
}
\\
\subfloat[$\rho_\infty$ for $z = 29$]{
\label{fig_pav_superficie_29-a}
\includegraphics[width=0.45\linewidth]{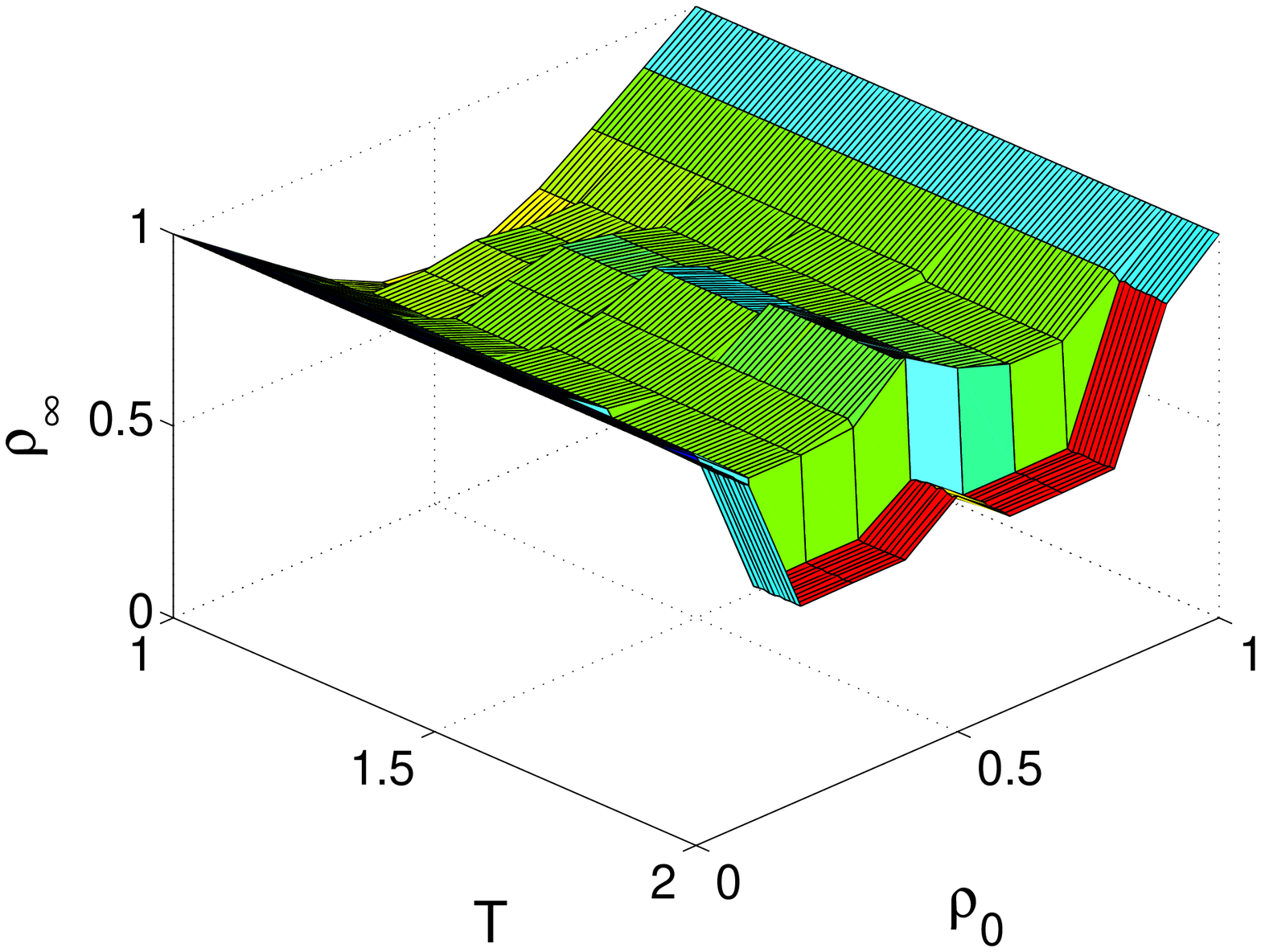}
}
\subfloat[$\rho_\infty$ standard deviation for $z = 29$.]{
\label{fig_pav_superficie_29-b}
\includegraphics[width=0.45\linewidth]{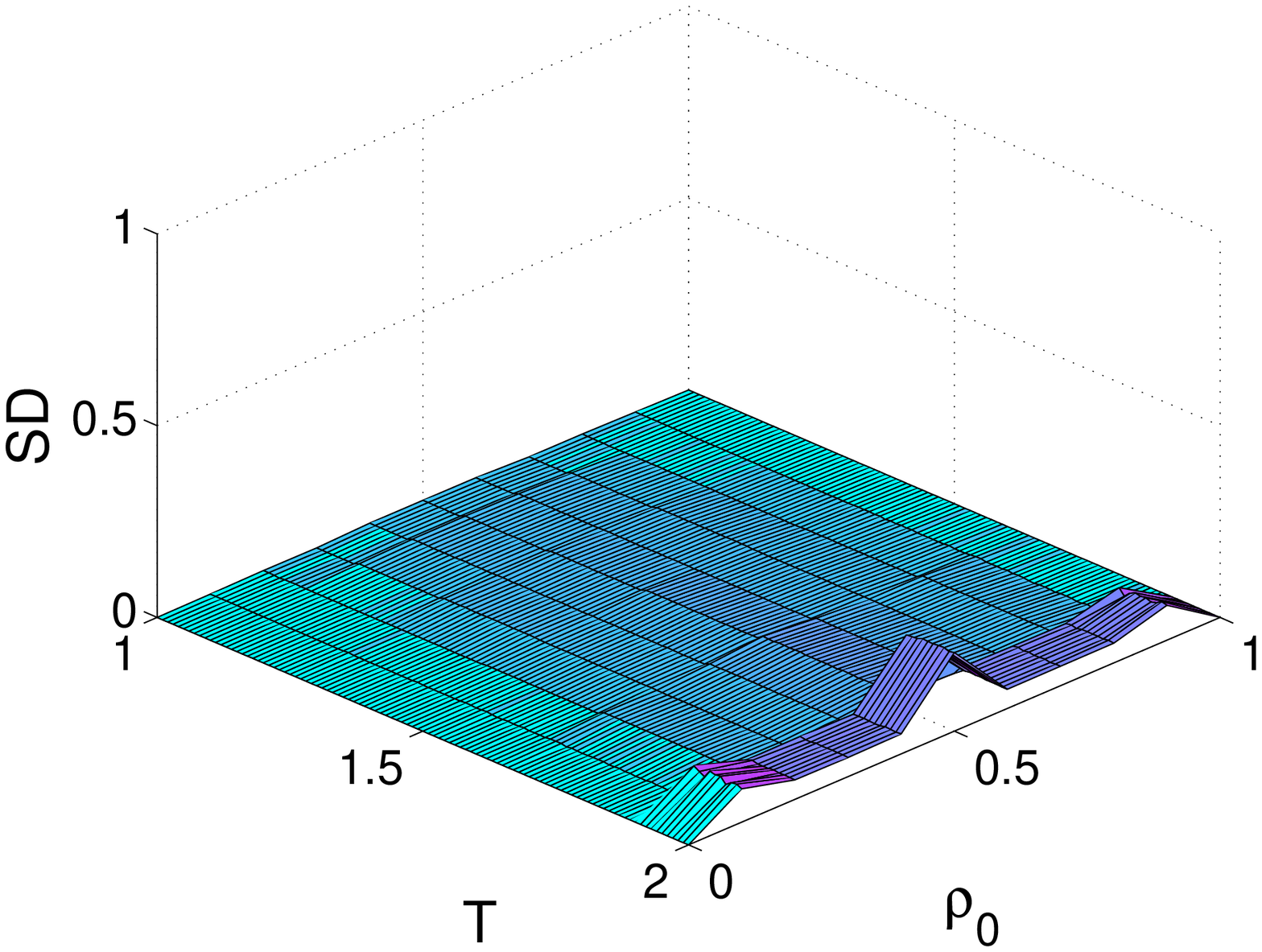}
}
\caption{
$\rho_\infty$ surface as function of the temptation value, $T$, and the initial cooperators temptation, $\rho_0$, for $z = \{3;9;29\}$.
}
\label{fig_pav_superficie_impar}
\end{figure}

To depict the asymptotic cooperators proportion in the steady state, we have used surfaces to show $\rho_\infty$ as a function of $T$ and $\rho_0$.
Figs.~\ref{fig_pav_superficie_par}~and
~\ref{fig_pav_superficie_impar} display these $\rho_\infty$ surfaces and their standard deviation for even (without self-interaction) and odd $z$ values (with self-interaction), respectively.
These phase-diagram present abrupt variations when the system passes through $T_c$, and, eventually, it may go from cooperative to {\it quasi}-regular phase.

An interesting aspect is the $\rho_\infty$ symmetry with respect to the $\rho_0 = 1/2$, that is, $\rho_\infty(\rho_0=1/2-\phi) = \rho_\infty(\rho_0=1/2+\phi)$ with $0 \leq \phi \leq 1/2$.
The self-interaction presence or absence, changes $T_c$ values and may change the phase (cooperative or {\it quasi}-regular) for the same region in the parameter space.
Notice that there are few regions where the standard deviation $SD$ is significant.

In the PES, if all players are cooperators ($\rho_0 = 1$), they always receive a positive payoff, and no player changes his/her state, therefore $\rho_\infty(T;~\rho_0=1;~z) = 1$.
Otherwise, if all players are defectors ($\rho_0 = 0$), in the first round, all players receive a negative payoff, and all of them switch their states, returning to the previous situation, consequently, $\rho_\infty(T;\rho_0=0;z) = 1$.
Thus, the $\rho_\infty$ symmetry around $\rho_0=1/2$ is a consequence of the PES.
For $\rho_0 = \beta$, $rN$ players receive a positive payoff and $(1-r)N$ players a negative payoff, whether, $\rho_0 = 1 - \beta$, $(1-r)N$ players receive positive payoff and $rN$ players negative payoff, generating the symmetry.

\begin{figure}[htbp]
\subfloat[$z=2$]{
\label{fig_pav_plots_T_z_02}
\includegraphics[width=0.45\linewidth]{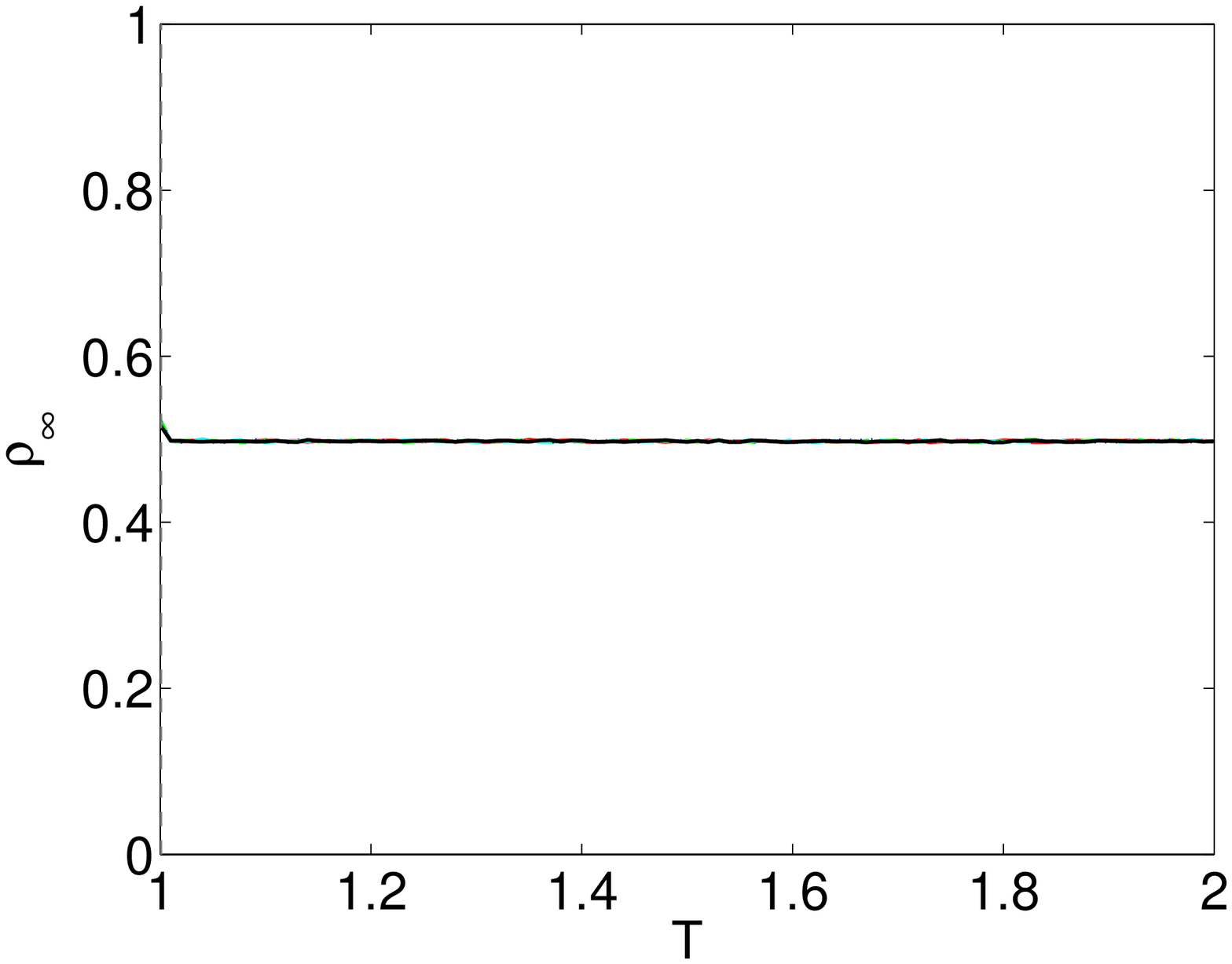}
}
\subfloat[$z = 3$]{
\label{fig_pav_plots_T_z_03}
\includegraphics[width=0.45\linewidth]{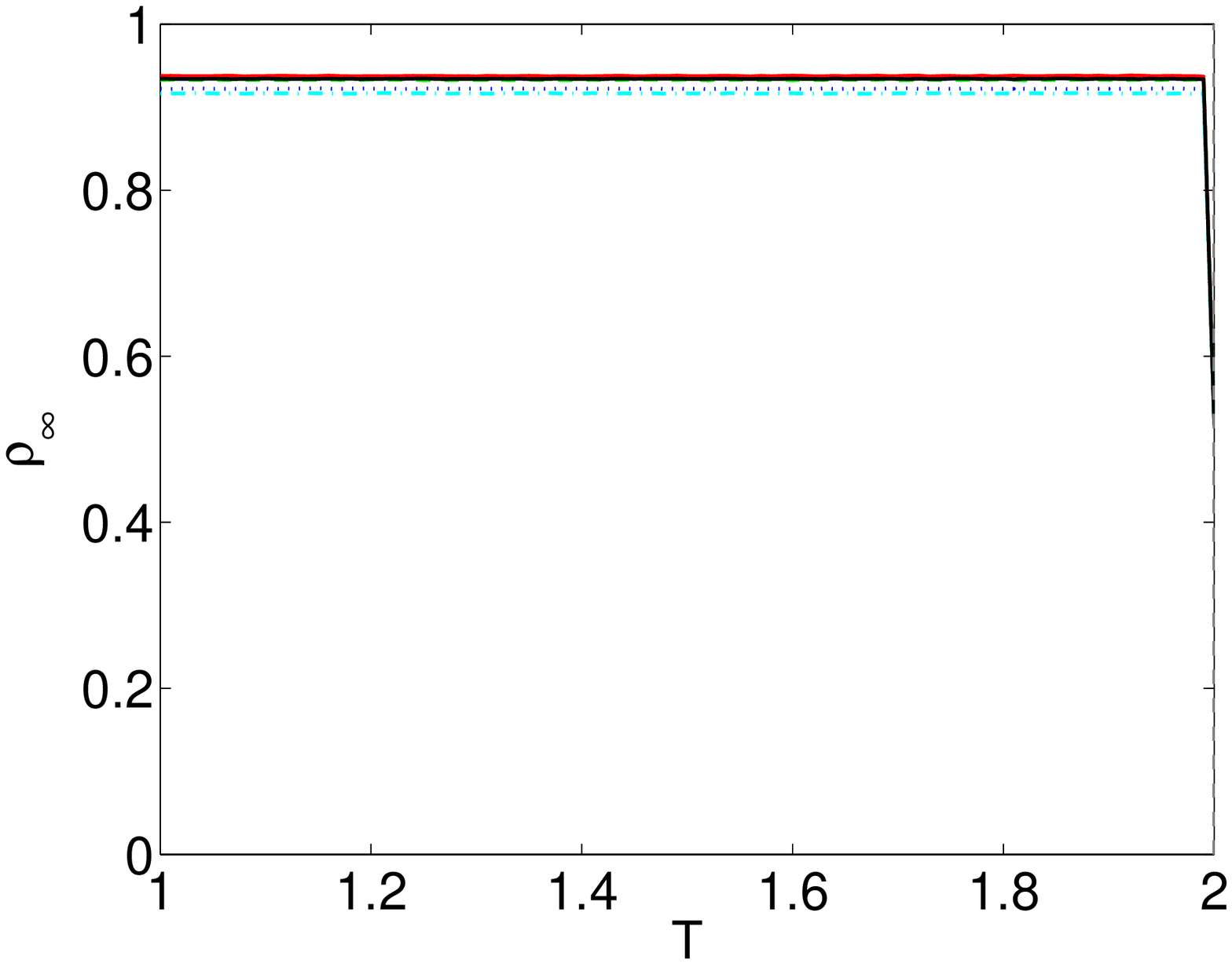}
}
\\
\subfloat[$z=8$]{
\label{fig_pav_plots_T_z_08}
\includegraphics[width=0.45\linewidth]{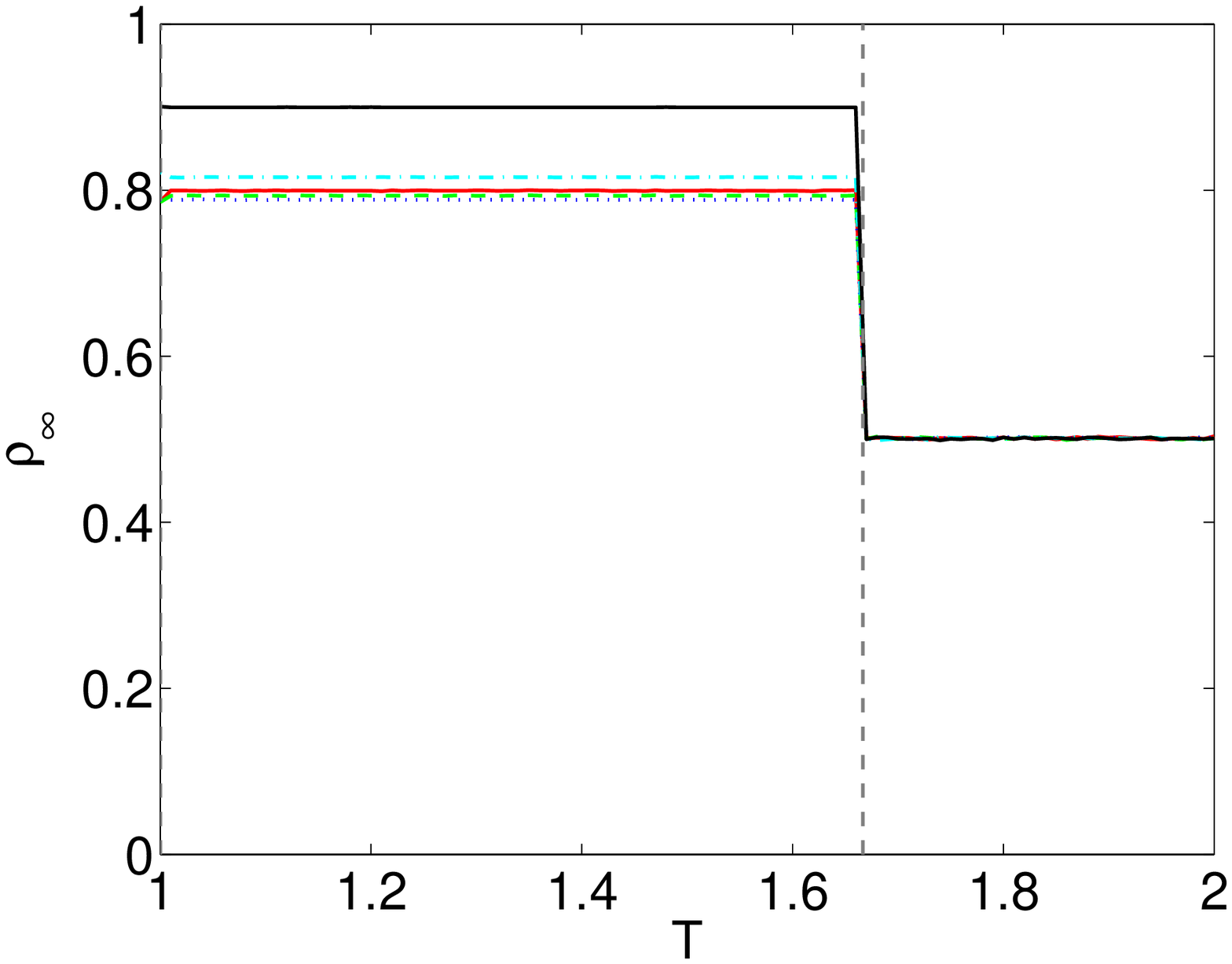}
}
\subfloat[$z=9$]{
\label{fig_pav_plots_T_z_09}
\includegraphics[width=0.45\linewidth]{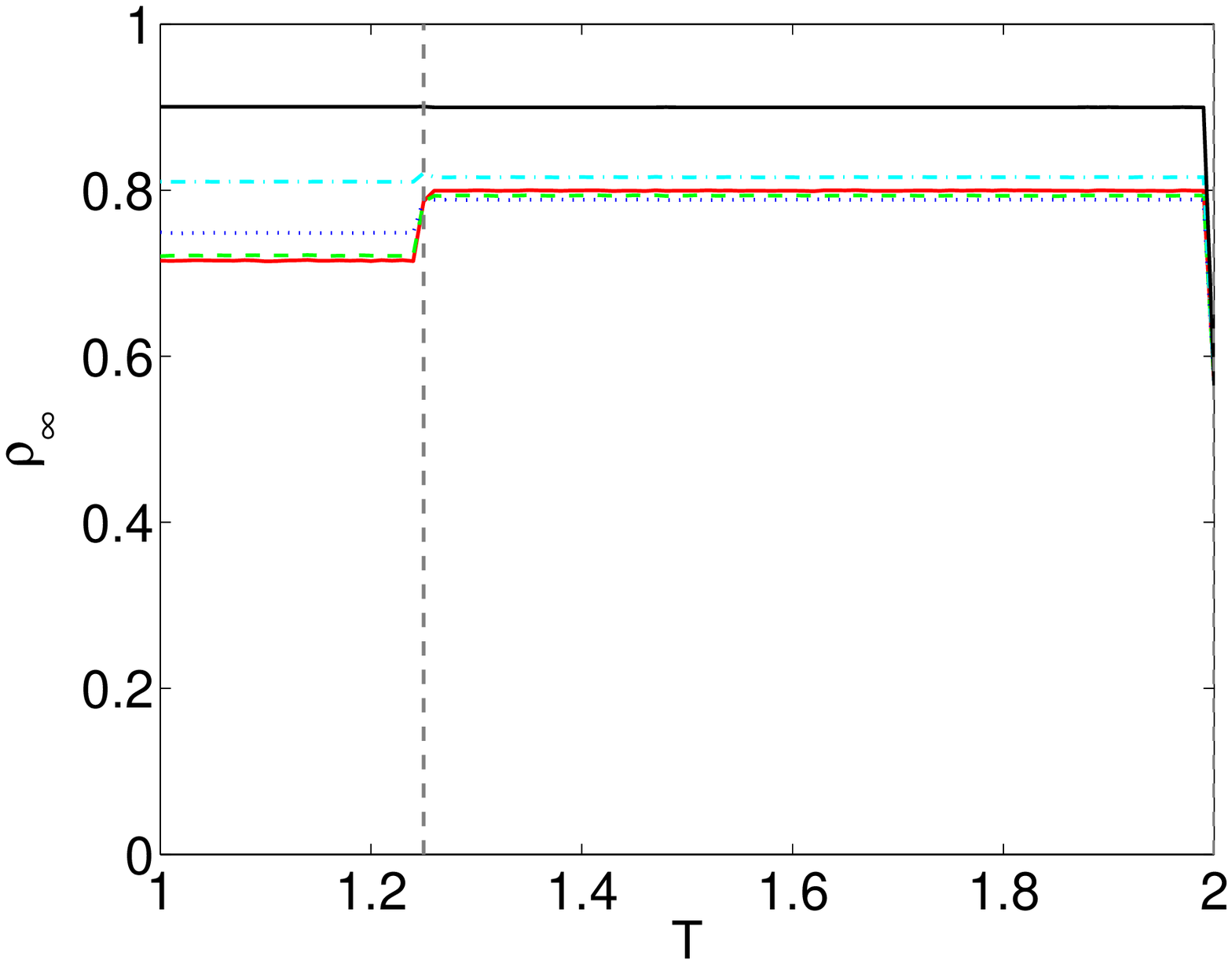}
}
\\
\subfloat[$z=28$]{
\label{fig_pav_plots_T_z_29}
\includegraphics[width=0.45\linewidth]{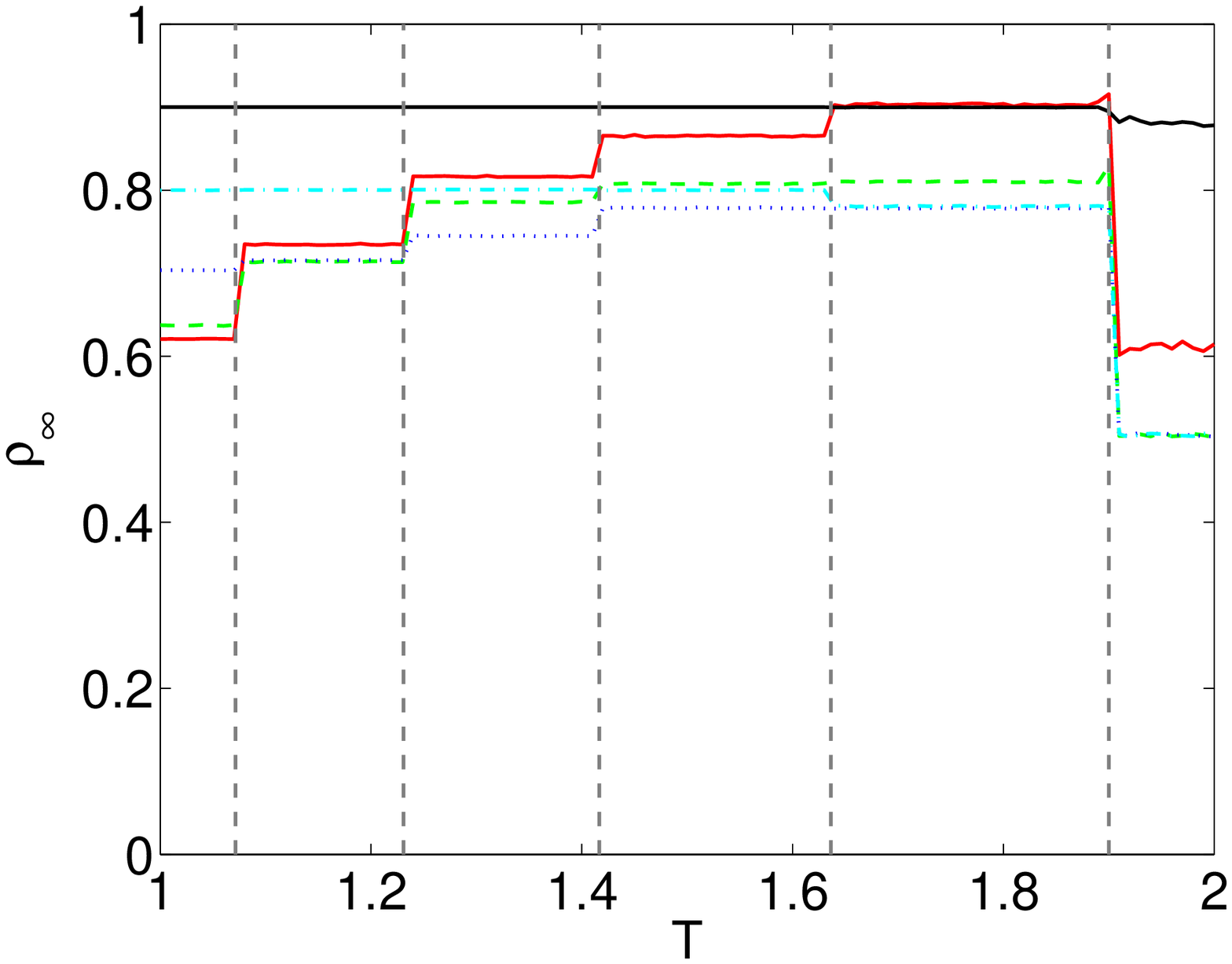}
}
\subfloat[$z=29$]{
\label{fig_pav_plots_T_z_30}
\includegraphics[width=0.45\linewidth]{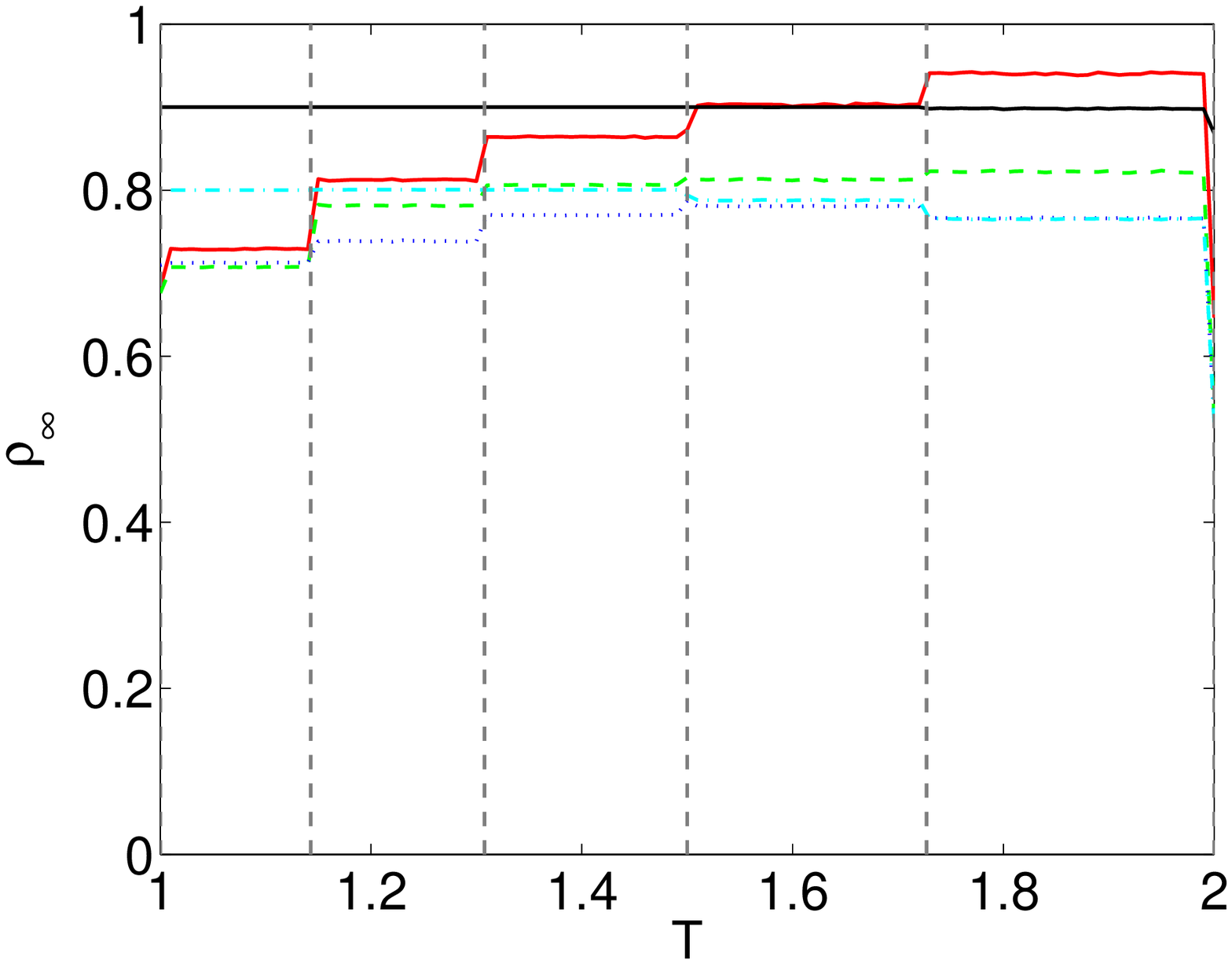}
}
\caption{
Asymptotic cooperator proportion ($\rho_\infty$) as function of temptation to defect $T$, for $z = \{2;~3;~8;~9;~29;~30\}$.
The vertical dashed lines sign $T_c$ in plots and these values are given by Eq.~\ref{eq_pavlov_Tc_generalizado}. 
Blue: $\rho_0=0.5$; green: $\rho_0=0.6$; red: $\rho_0=0.7$; cyan: $\rho_0=0.8$; magenta: $\rho_0=0.9$.
}
\label{fig_pav_plots_T_z}
\end{figure}

The surface projection $\rho_\infty(T,\rho_0,z)$ at plane $\rho_\infty T$ shows $\rho_\infty$ as function of $T$ for different $\rho_0$ values.
In Figure~\ref{fig_pav_plots_T_z}, one sees the plots for some even and odd $z$ values.
The transitions in $\rho_\infty$ can be seen when the parameter $T$ passes through critical temptation thresholds, $T_c$, given by Eq.~\ref{eq_pavlov_Tc_generalizado}. 
In the plots, the $T_c$ values marked by dashed vertical lines. 
For example, in Fig.~\ref{fig_pav_plots_T_z_08} ($z=8$ - without self-interaction) $T_c = 5/3$.
Meanwhile, in Fig.~\ref{fig_pav_plots_T_z_09} ($z=9$ - with self-interaction) $T_c = \{5/4;~2\}$.

The results show that the cooperative phase is more prominent than {\it quasi}-regular one.
Increasing the $z$ values, the quantity of $T_c$ values raises.
When the system goes through $T_c$, $\rho_\infty$ varies.
The non dependence on a group, provides more liberty to each player seek the best outcome that satisfy his/her own aspiration level.
In this case, the worst result is the homogeneity among the players payoff ({\it quasi}-regular phase), instead of cooperation, the best result, but much better than the defection.

\subsection{Spatio-temporal patterns} 
\label{resultados_pavlov_padroes}

The patterns formed by cooperative/defective clusters that emerge in the system are yielded from local interactions among players.
When the players adopt the DES, the differences between the border players payoff are fundamental to determine the system dynamics~\cite{pereira_2008_IJMPC}.
While in the PES, these border payoff differences are not as crucial as in the DES.
For a more detailed explanation of patterns formation, see the Appendix~\ref{app_pattern}.
The spatio-temporal representation show us that these clusters can form {\it fingers} and {\it gliders} and they can interact.

One may notice that clusters composed exclusively by cooperators sustain itself by the maintenance of cooperation among them.
However, cooperation remains only when the size of the cooperative cluster is large enough to maintain a positive payoff to its members.
The members placed in the borders are exploited by defectors, but they do not switch their states because their payoffs are positive, despite they are lower than the payoff of exploiters and inner players of the cooperative cluster.

If a defective cluster is large enough to produce negative payoffs to their members, it will not be stable.
Thus, players with negative payoffs, will switch their states immediately.
Thereby, Tragedy of Commons\footnote{The Tragedy of Commons occurs when multiple individuals act independently, aiming only his own interest.
When this action is done by multiple individuals simultaneously, it can destroy the advantage desired by all of them, for example, finishing the desired resources in the environment.
However it is not the purpose of anyone.} does not occur, because the players negative payoff does not persist for more than one round, such as for players adopting the DES.
In this way, the population mean payoff is higher adopting the PES than the DES, where the Tragedy of Commons takes longer to vanish (when it vanishes)~\cite{pereira_2008_IJMPC,pereira_2008_BJP}.
In the PES, a player uses his/her own payoff to decide whether he/she will switch his/her state or not.
It is an individual decision based on the personal aspiration level.
Therefore, collective features, i.e. players switching their states due the environment (group) where they are inserted, may not occur.

Different neighborhood configurations may generate positive or negative payoffs for the players.
It is simple to calculate the maximum defective cluster size and the minimum cooperative cluster size, which can remain together during the system evolution (stable clusters).
In the cooperative cluster case, player $i$ does not switch his/her state, if there are at least $c_{\min}$ cooperators in his/her neighborhood.
This guarantees a positive payoff $G_i^{c_{\min}}(\theta_i = 1) > 0$, thus, from Eq. \ref{eq_ganho_total_generalizado} one has:
\begin{equation}
	c_{\min} > z \left(\frac{1}{1+R/T}\right).
\end{equation}
The situation is reversed in the defective cluster case.
In this case, the player $i$ must have in his/her neighborhood a maximum of $d_{\max}$ defectors, so that $G_i^{d_{\max}} (\theta_i = 0) > 0$, and:
\begin{eqnarray}
	d_{\max} < z \left(\frac{1}{1+R/T}\right).
\end{eqnarray}

A {\it finger} is a cluster that extends itself along a straight line during time evolution.
It can be simple (flat one), or complex (composed by regular oscillations, like a saw-tooth, for example).
The {\it finger} interior can be composed by cooperators/defectors or by intricated combinations of cooperators and defectors.
It may present symmetry with respect to central player of the pattern and periodicity in the player states.
A {\it glider} is a cluster that extends itself diagonally, and it has the same features of the {\it fingers}.

In short, defective {\it fingers} may be composed of at most $d_{\max}$ defectors, and cooperative ones of at least $c_{\min}$ cooperators.
For instance, for $z = 2$, {\it fingers} formed by up to two players are always smooth and continuous (see simple and complex {\it fingers} in the Appendix~\ref{app_pattern}).
In general, the stable clusters are the cooperative ones (with at least $c_{\min}$ cooperators) and small defective ones.
Defective clusters that are greater than $d_{\max}$ destabilize themselves rapidly in few rounds.

\begin{figure}[!htbp]
\centering
\subfloat[]{\label{fig_pav_arvore-a}\includegraphics[width=0.45\linewidth]{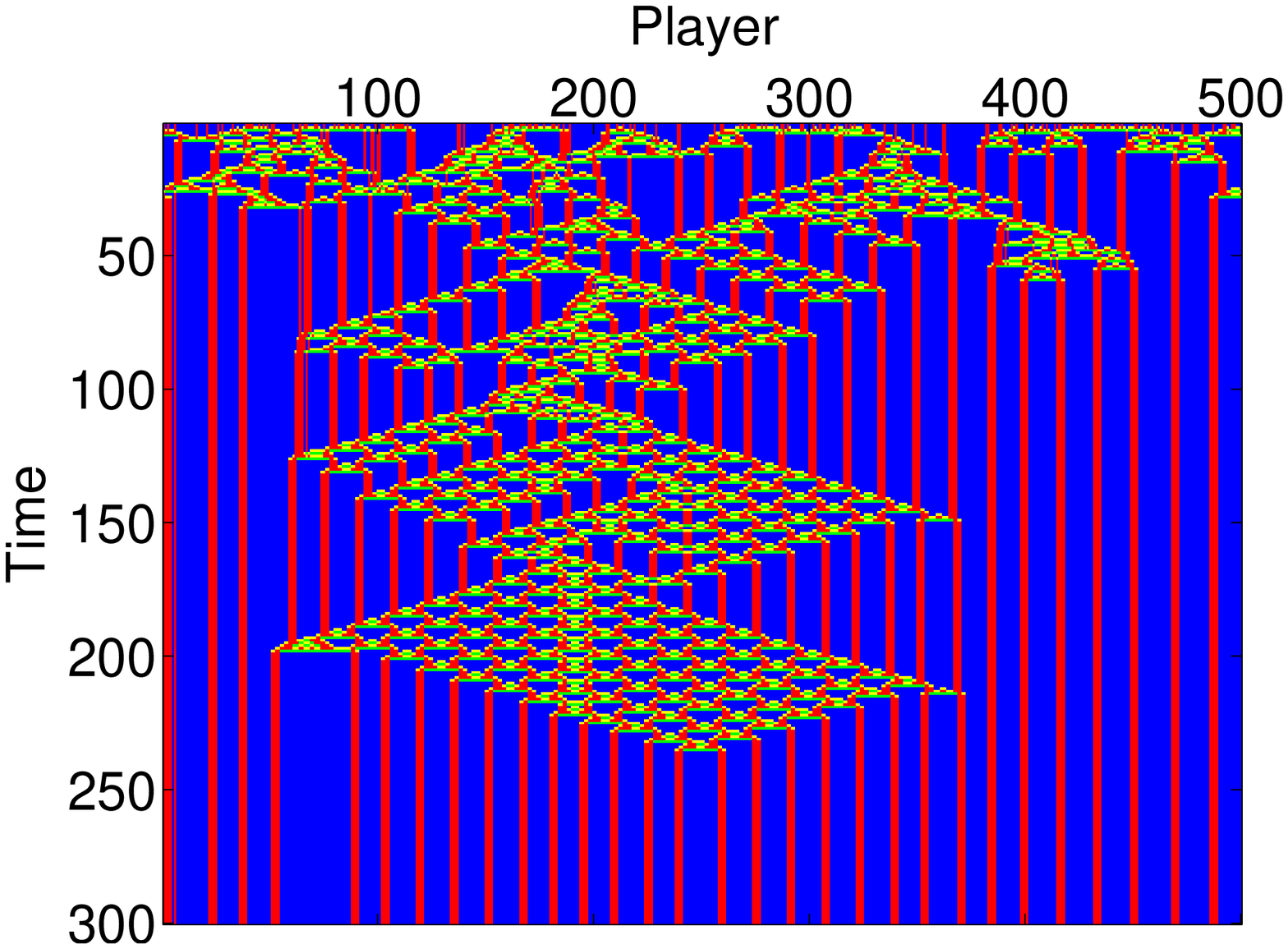}}
\subfloat[]{\label{fig_pav_arvore-b}\includegraphics[width=0.45\linewidth]{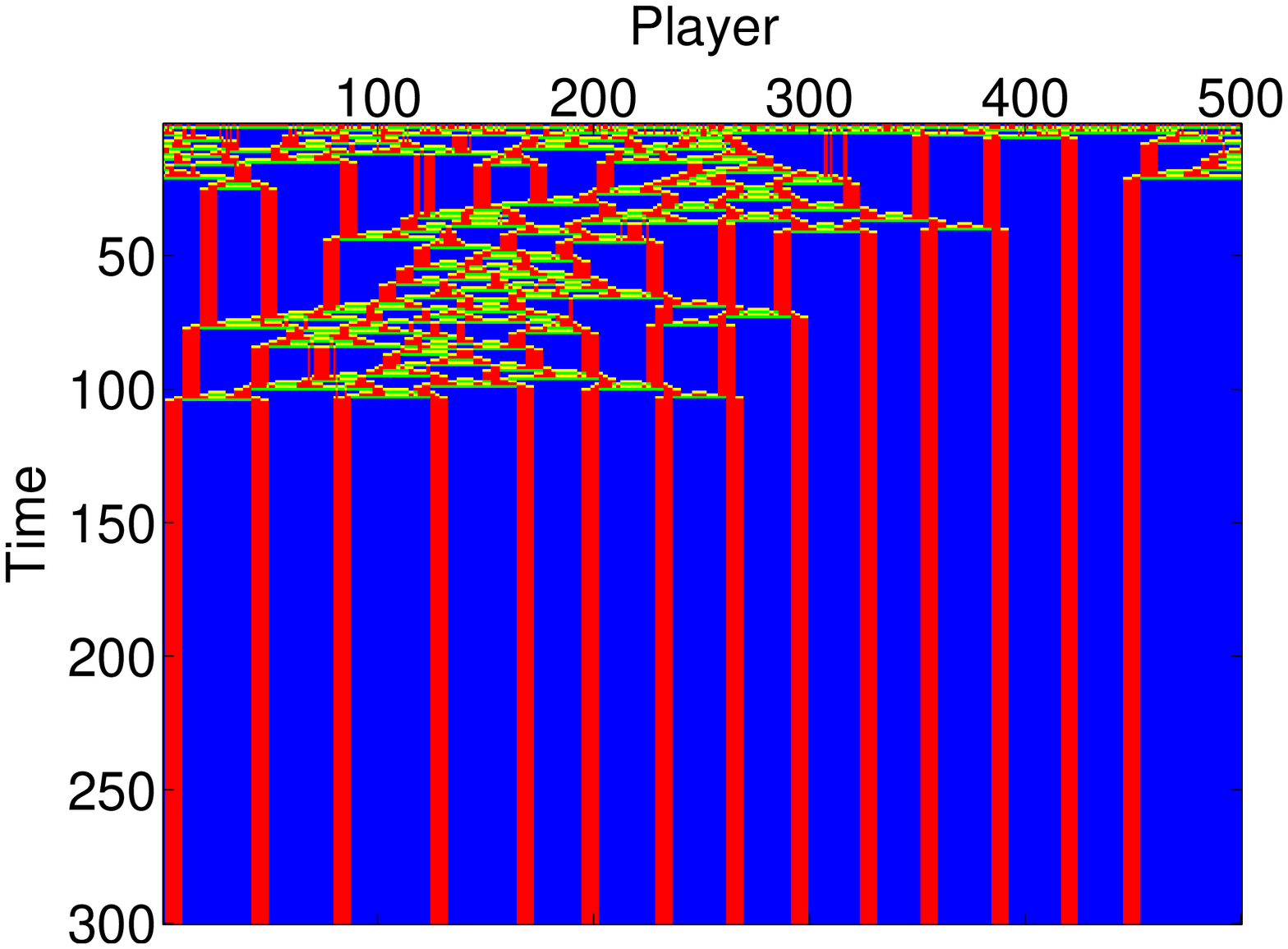}}
\\
\subfloat[]{\label{fig_pav_arvore-c}\includegraphics[width=0.45\linewidth]{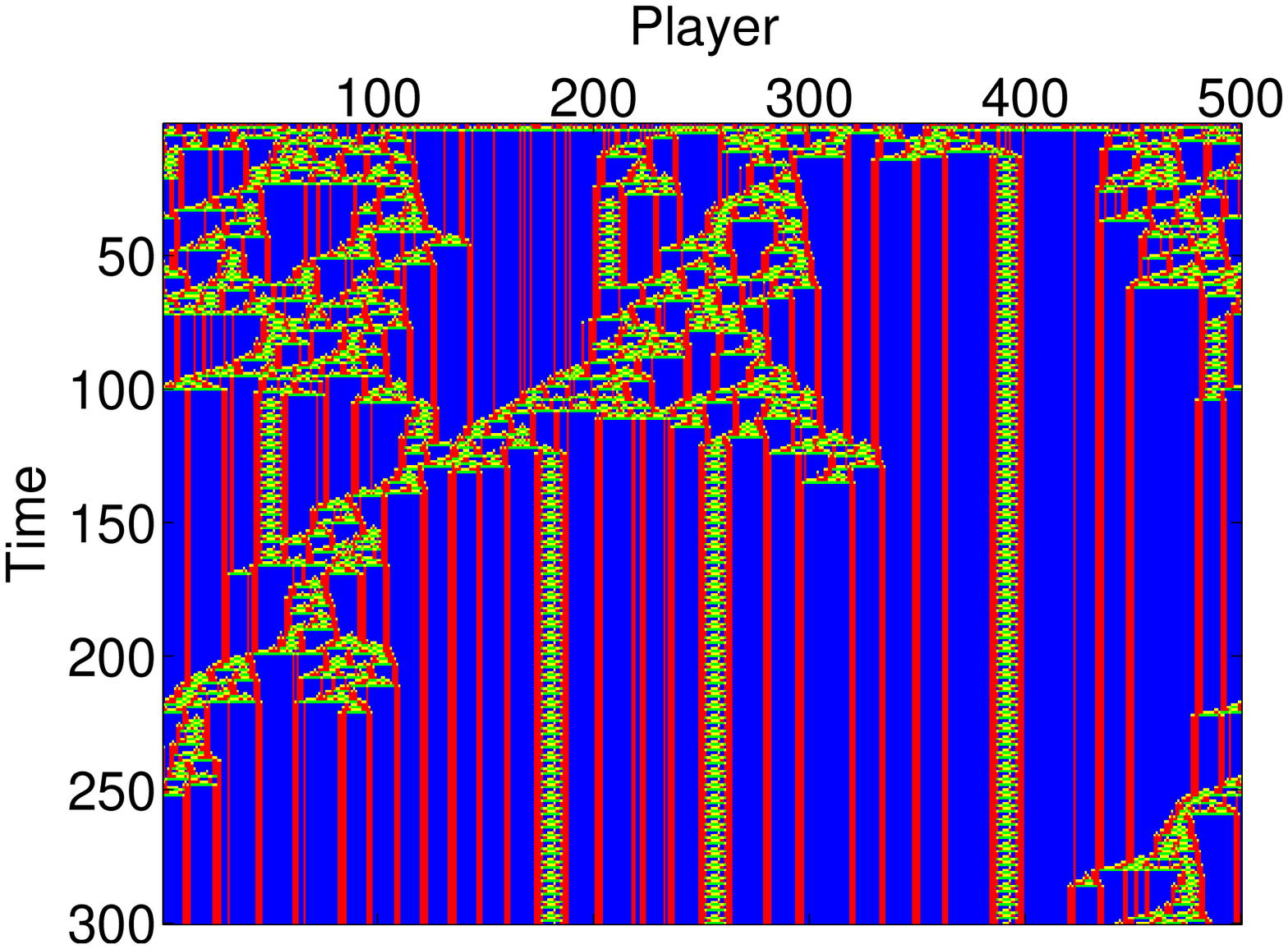}}
\subfloat[]{\label{fig_pav_arvore-d}\includegraphics[width=0.45\linewidth]{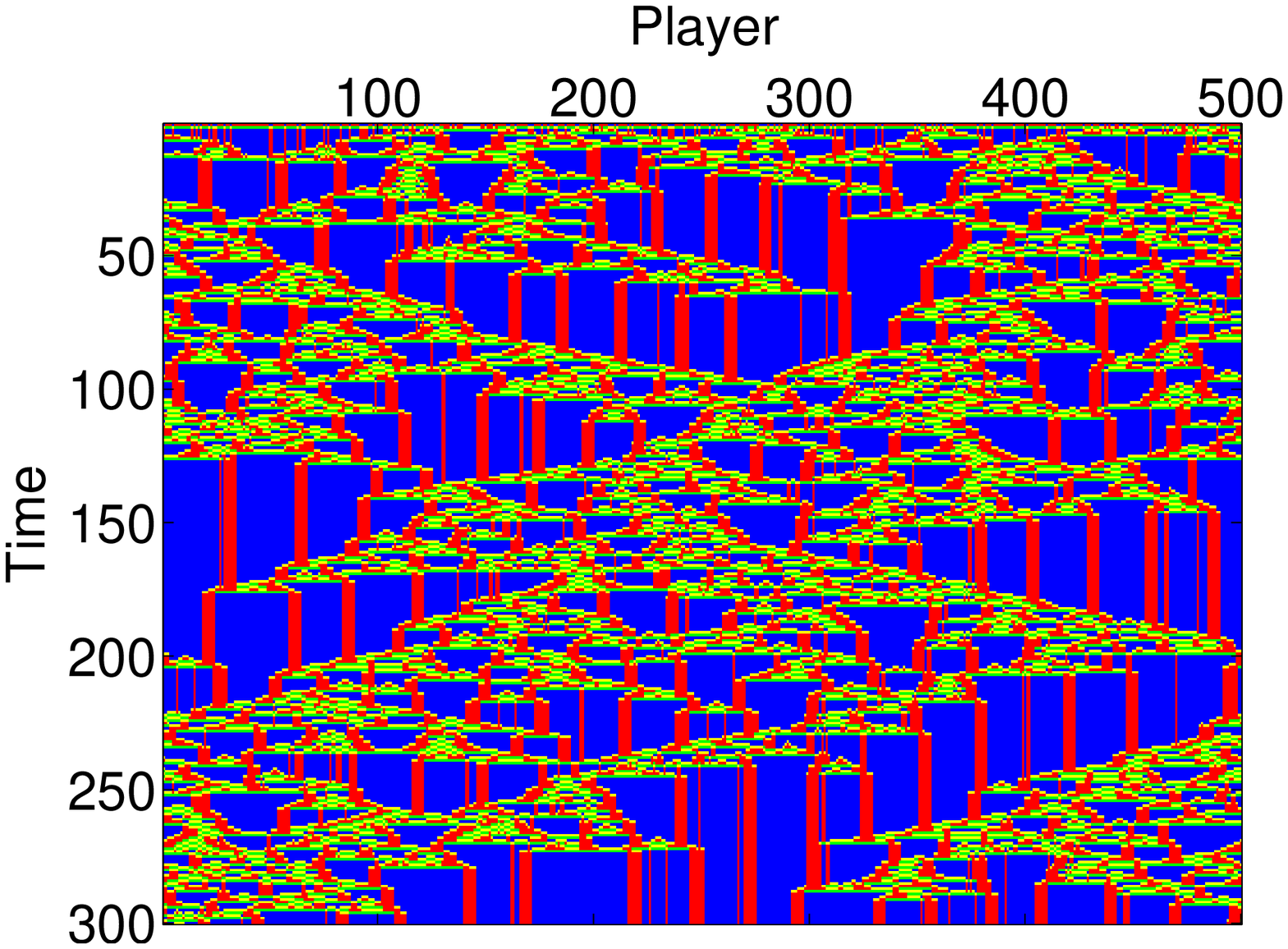}}
\caption{
Illustration of intersections of {\it glider} with {\it fingers} and others {\it glider}.
The parameters of these simulations are:
$L=500$, $t=300$ and 
(a) $T=1.40$, $\rho_0=0.7$, $z=14$ (without self-interaction); 
(b) $T=1.70$, $\rho_0=0.3$, $z=26$ (without self-interaction); 
(c) $T=1.40$, $\rho_0=0.5$, $z=12$ (without self-interaction); and 
(d) $T=1.40$, $\rho_0=0.3$, $z=24$ (without self-interaction).
}
\label{fig_pav_arvore}
\end{figure}

The transient regime is the necessary time to cease the patterns interactions or stabilize the patterns propagation and it varies depending on the parameters set used (see Fig.~\ref{fig_pav_arvore}).
Other possibility is the emergence of the {\it quasi}-regular phase, which is stationary, but there is a very large number of players who switch their states, but do not emerge {\it fingers} or {\it gliders}, and $\rho_\infty \sim 0.5$.

The intersection among cluster patterns generates very interesting structures.
For example, Fig.~\ref{fig_pav_arvore} illustrates the presence of {\it gliders}\footnote{In systems that adopt the DES, the inclination of {\it glider} is determined by the direction of upgrade of players states.
On the other hand, if the system adopts the PES, the {\it glider} can propagate both from left to right-hand side or {\it vice-versa}.}
that interact among themselves and with {\it fingers}.
These interactions can generate either simple (Fig.~\ref{fig_pav_arvore-a}-\ref{fig_pav_arvore-d}) or complex (Fig.~\ref{fig_pav_arvore-c}) {\it fingers}.

One can understand the {\it quasi}-regular phase by observing the cooperative/defective clusters behavior.
If defectors of a particular cluster receive a negative payoff at moment $t$, these players switch their status to cooperators at $t + 1$.
If this action is synchronized among different clusters, cooperation may emerge.
Otherwise, if they are not synchronized and one cluster switches at the instant $t$ and its neighbors at the instant $t + 1$ these clusters alternate between cooperation and defection, and there is a balance among cooperators that switch their states to defectors and {\it vice-versa} keeping the proportion of cooperators almost constant, with small oscillations.

\begin{figure}[htbp]
\centering
\subfloat[]{\label{fig_pav_quasi_1-a}\includegraphics[width=0.8\linewidth]{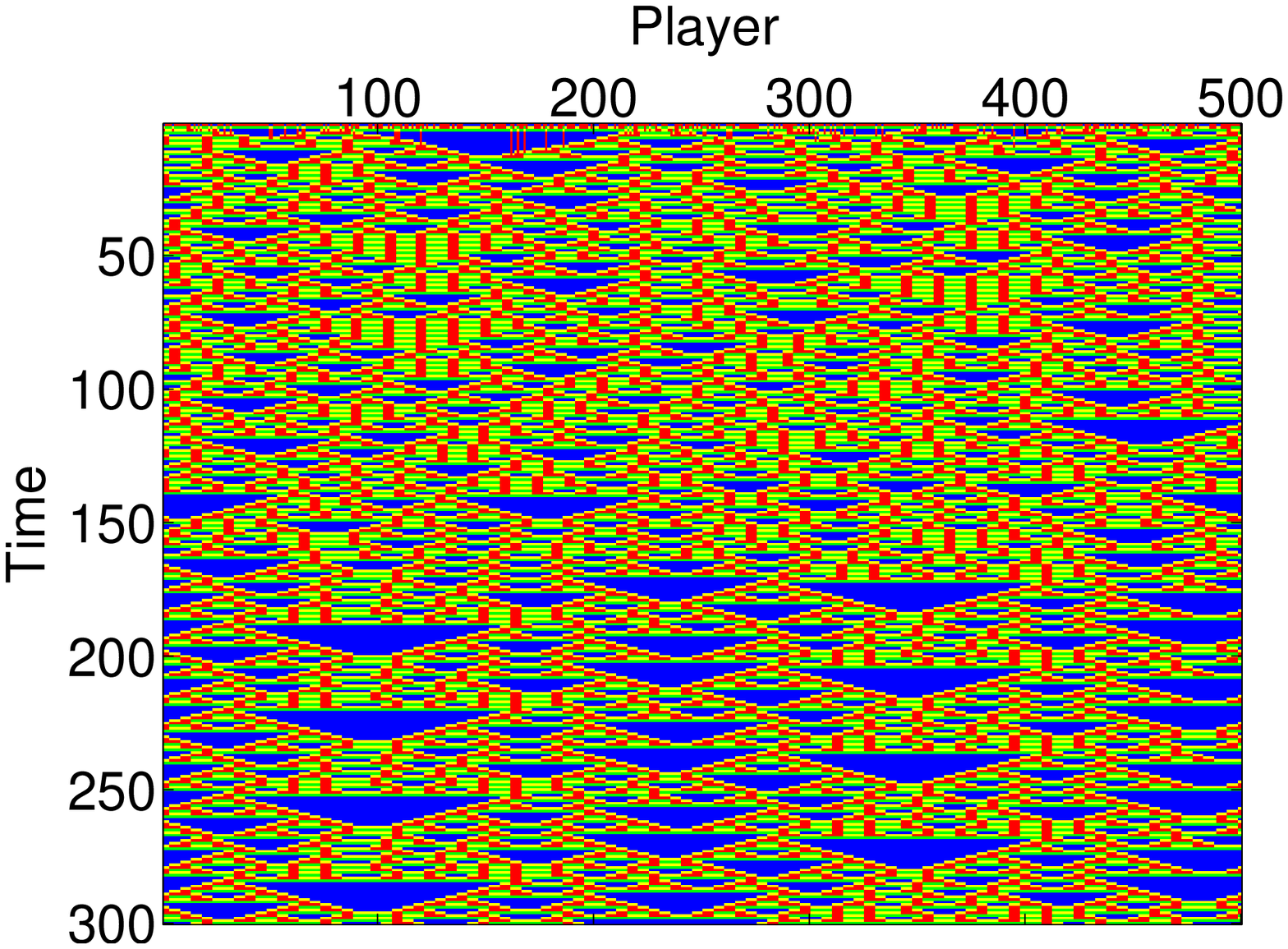}}
\\
\subfloat[]{\label{fig_pav_quasi_1-b}\includegraphics[width=0.8\linewidth]{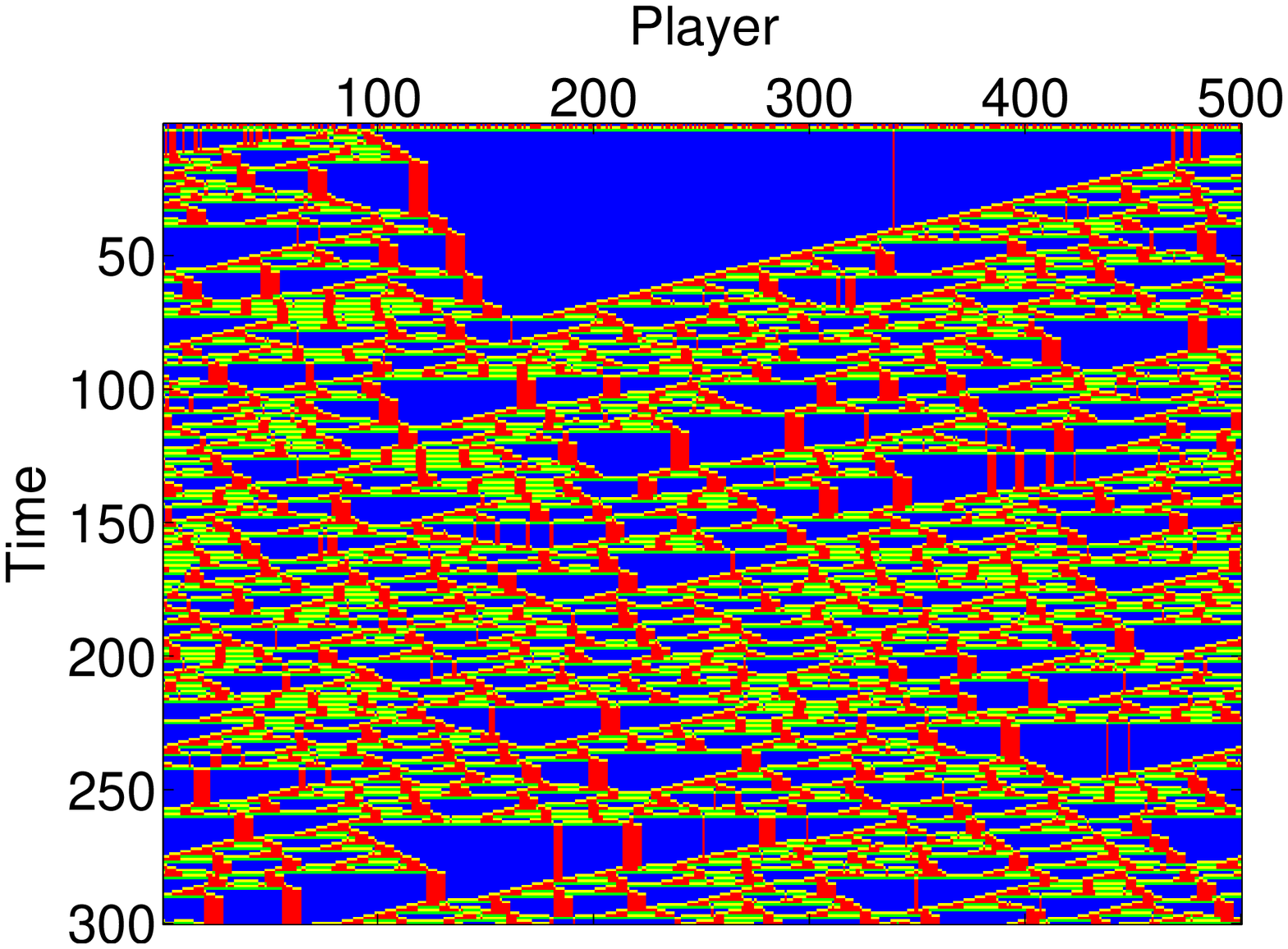}}
\caption{
Formation and evolution of the {\it quasi}-regular system.
The parameters of these simulations are: $L=500$, $t=300$ and
(a) $T=1.90$, $\rho_0=0.3$, e $z=26$ (without self-interaction) and 
(b) $T=1.90$, $\rho_0=0.5$, and $z=29$ (with self-interaction).
Note the presence of a triangle-like at time $t = 50$ at players 200-250.}
\label{fig_pav_quasi_1}
\end{figure}

\begin{figure}[htbp]
\centering
\includegraphics[width=1.0\linewidth]{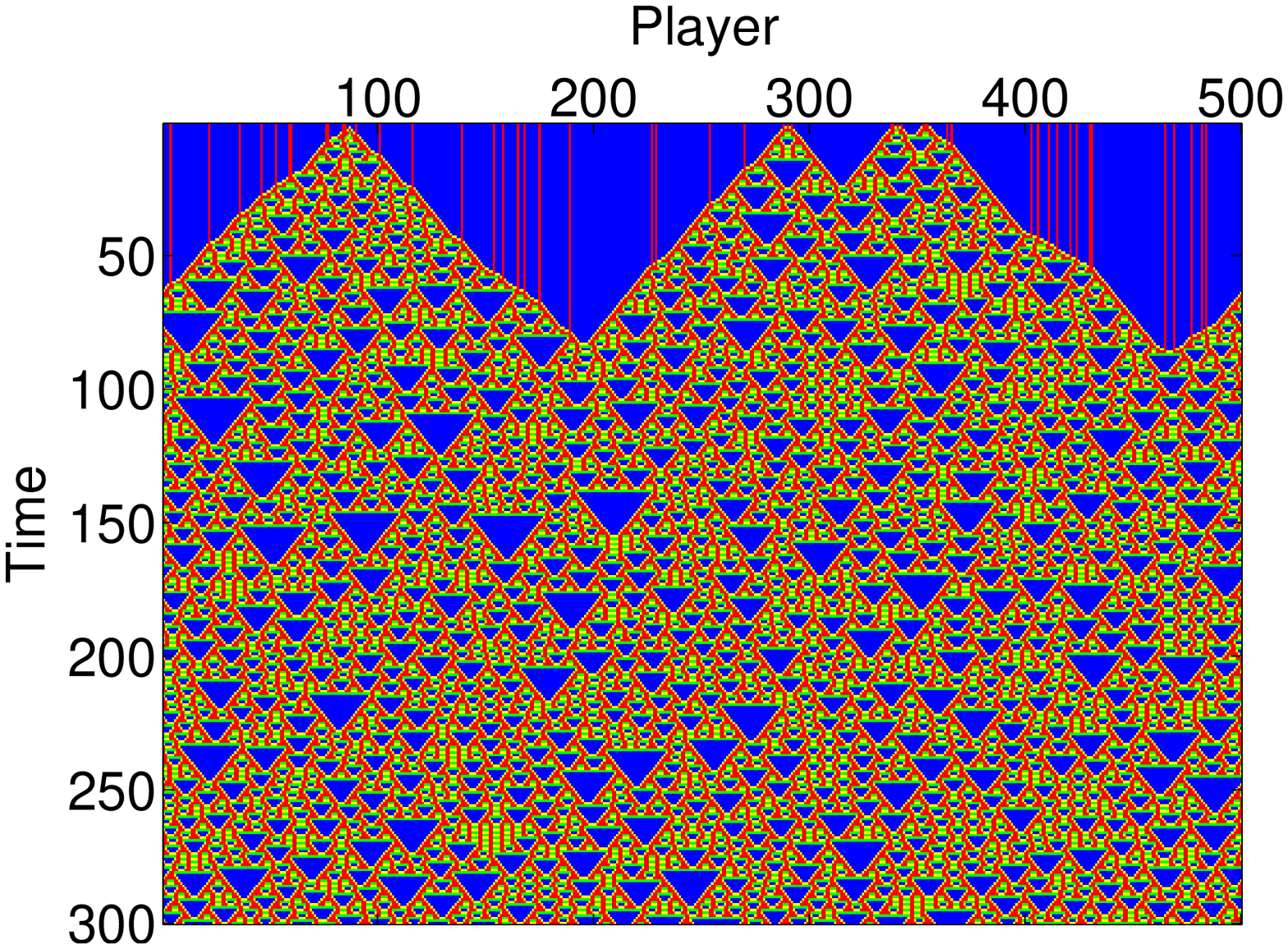}
\subfloat[]{\label{fig_pav_quasi_2-a}\includegraphics[width=0.45\linewidth]{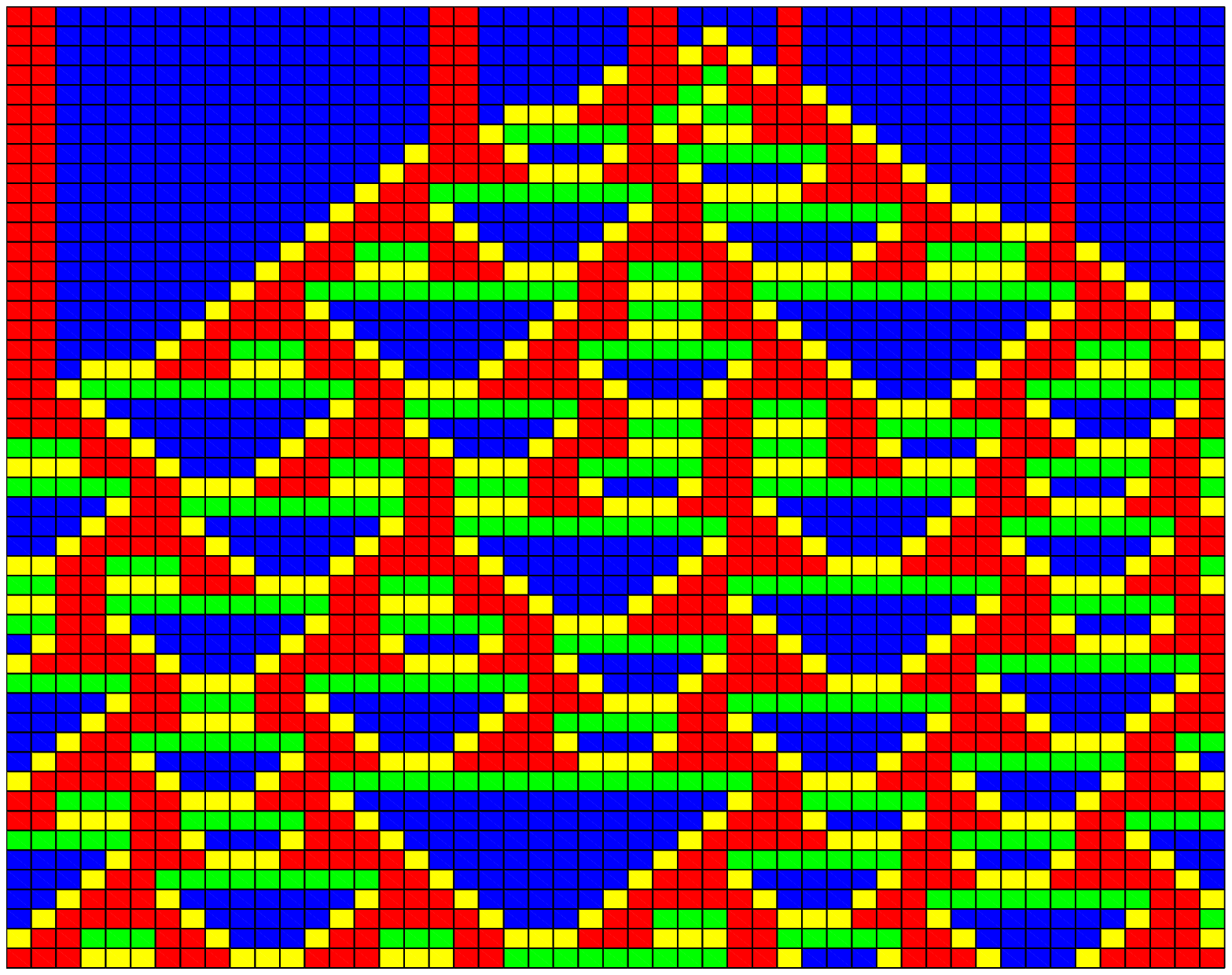}}
\subfloat[]{\label{fig_pav_quasi_2-b}\includegraphics[width=0.45\linewidth]{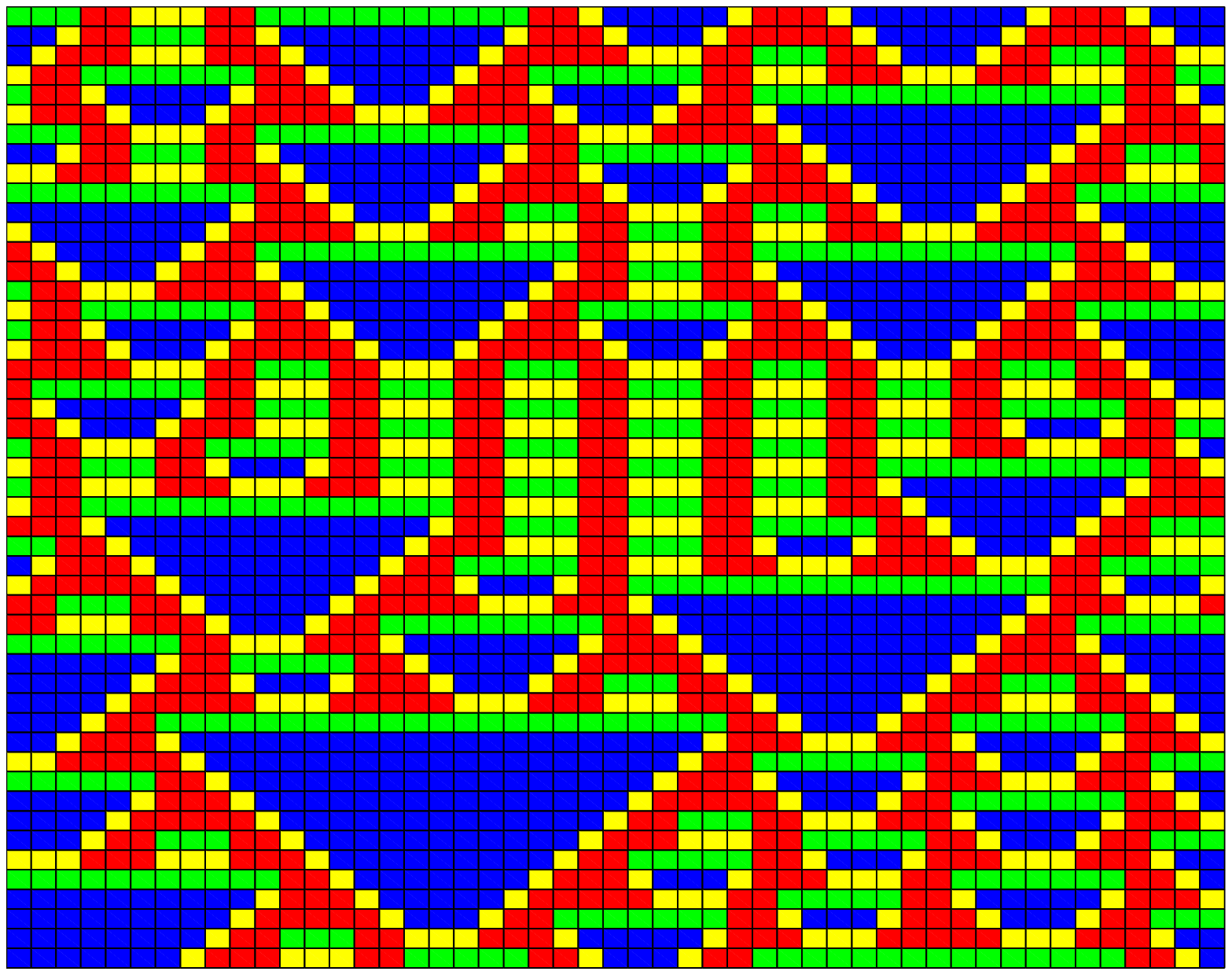}}
\caption{
Formation and evolution of the {\it quasi}-regular system.
The parameters of this simulation are:
$L=500$, $t=300$, $T=2.00$, $\rho_0=0.9$ and $z=6$ (without self-interaction).
(a) zoom of the region around to $t = 1$ for the players close to the player $90$;
(b) zoom of the region: $t = 260$ around player $160$.}
\label{fig_pav_quasi_2}
\end{figure}

In Figs.~\ref{fig_pav_quasi_1}-\ref{fig_pav_quasi_2}, one sees some examples where the synchronization among clusters had not occurred, $\rho_\infty \sim 0.5$ and many players switch their states at each round, giving rise to the {\it quasi}-regular phase.
The triangles appear when adjacent defectors switch their states to cooperation at the same time.
In Fig.~\ref{fig_pav_quasi_1-a}, a transient followed by the {\it quasi}-regular phase with periodicity is presented.
It also appears another triangular pattern, but its interior is not composed exclusively by cooperators or defectors, but by complex cooperative and defective {\it fingers}.
This pattern is a triangle, with sides not well defined, which we call triangle-like, see Fig.~\ref{fig_pav_quasi_1-a}.
In Fig.~\ref{fig_pav_quasi_1-b}, one can notice that the cooperative clusters size are larger for system with higher connectivity, but the system phase remains {\it quasi}-regular.

In Fig.~\ref{fig_pav_quasi_2}, one sees that few defective clusters are enough to drive the system to {\it quasi}-regular phase instead of cooperative one, despite the high initial cooperators proportion, this occurs because $T = 2$.
In Fig.~\ref{fig_pav_quasi_2-a}, one can see a defective clusters zoom.
Fig.~\ref{fig_pav_quasi_2-b} illustrates a so-called triangle-like that emerges at $t = 260$ around player $160$.

\begin{figure}[!htbp]
\centering
\subfloat[$T=1.00$]{\includegraphics[width=0.45\linewidth]{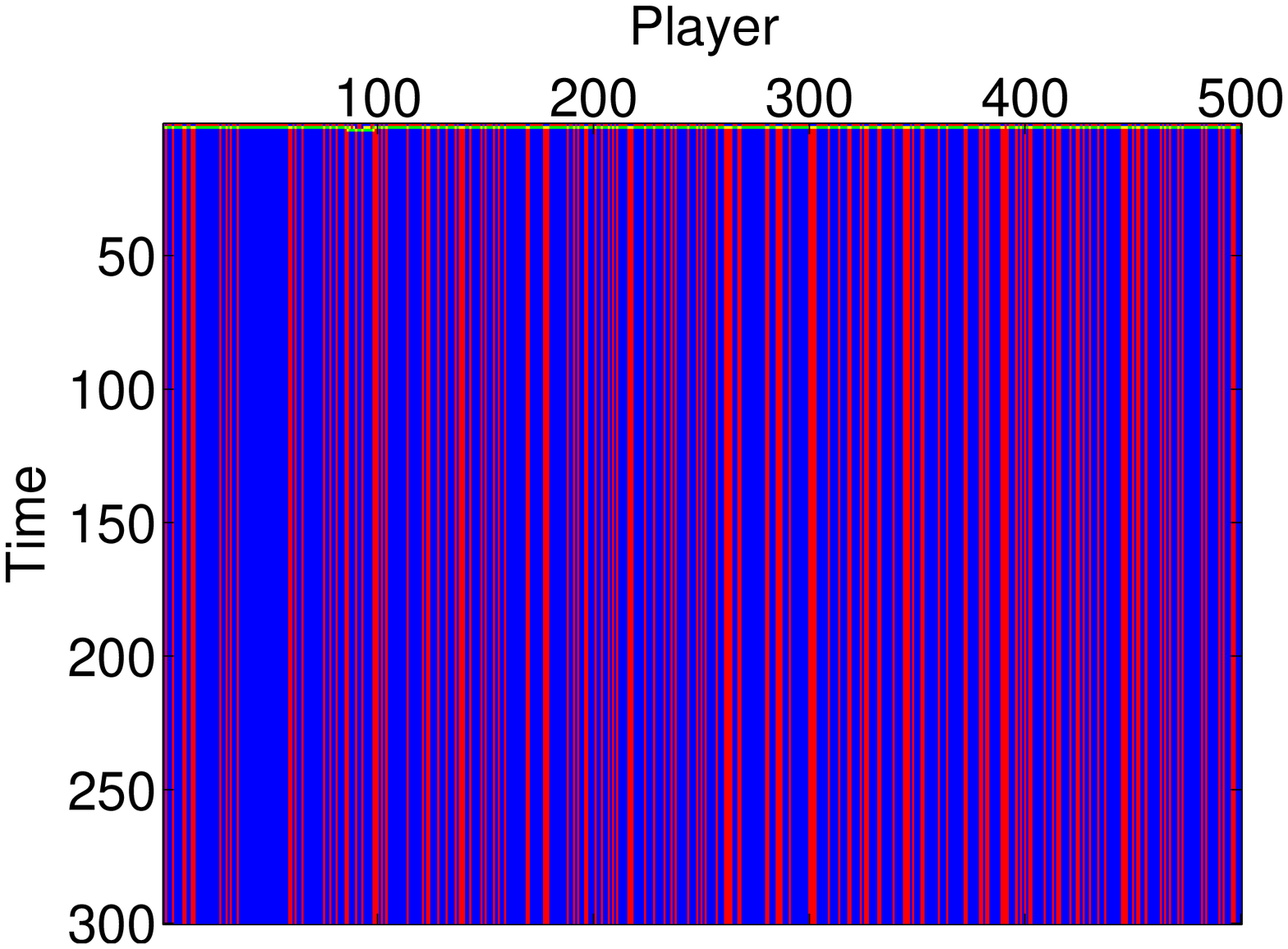}}
\subfloat[$T=1.01$]{\includegraphics[width=0.45\linewidth]{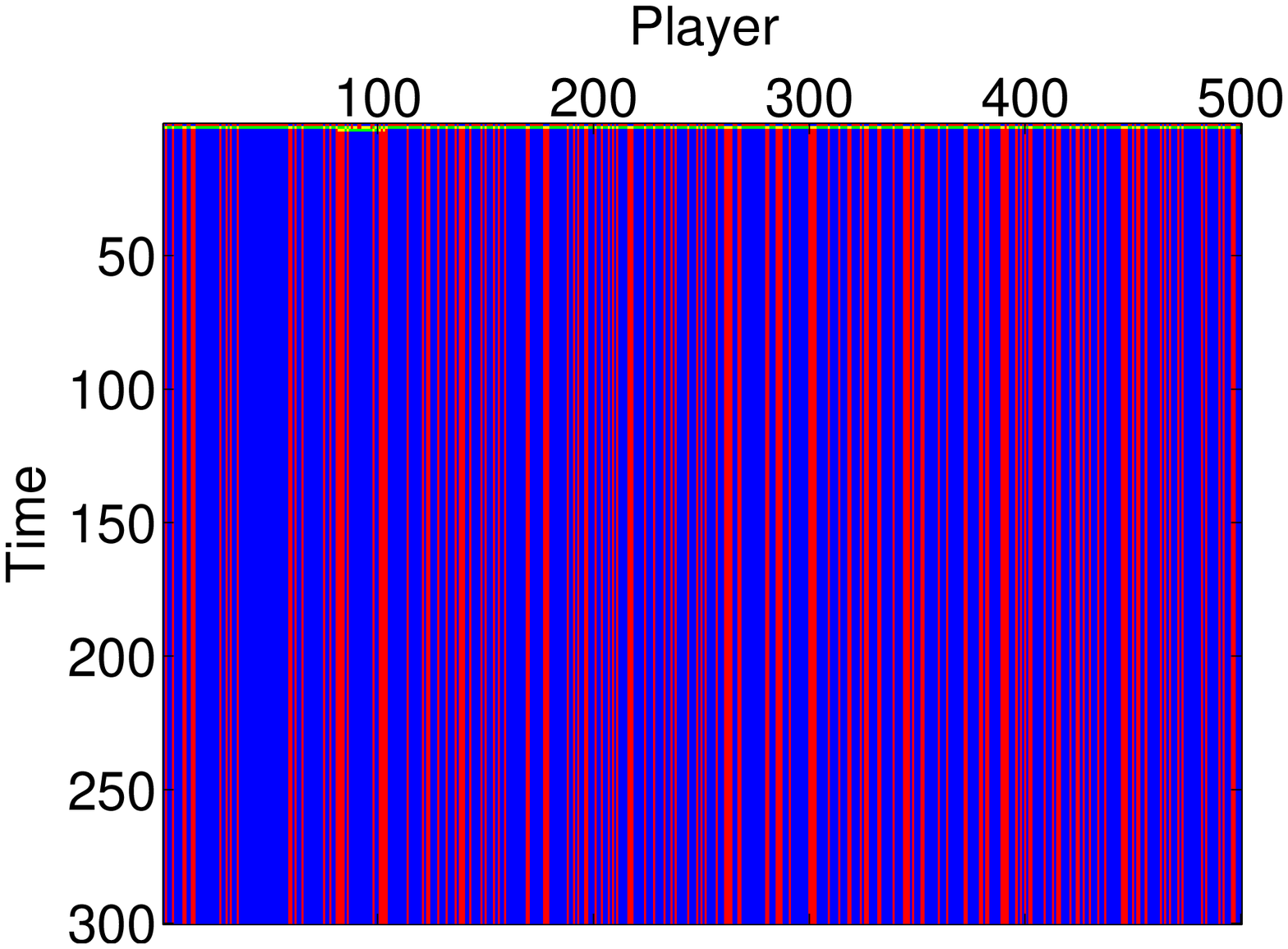}}
\\
\subfloat[$T=1.19$]{\includegraphics[width=0.45\linewidth]{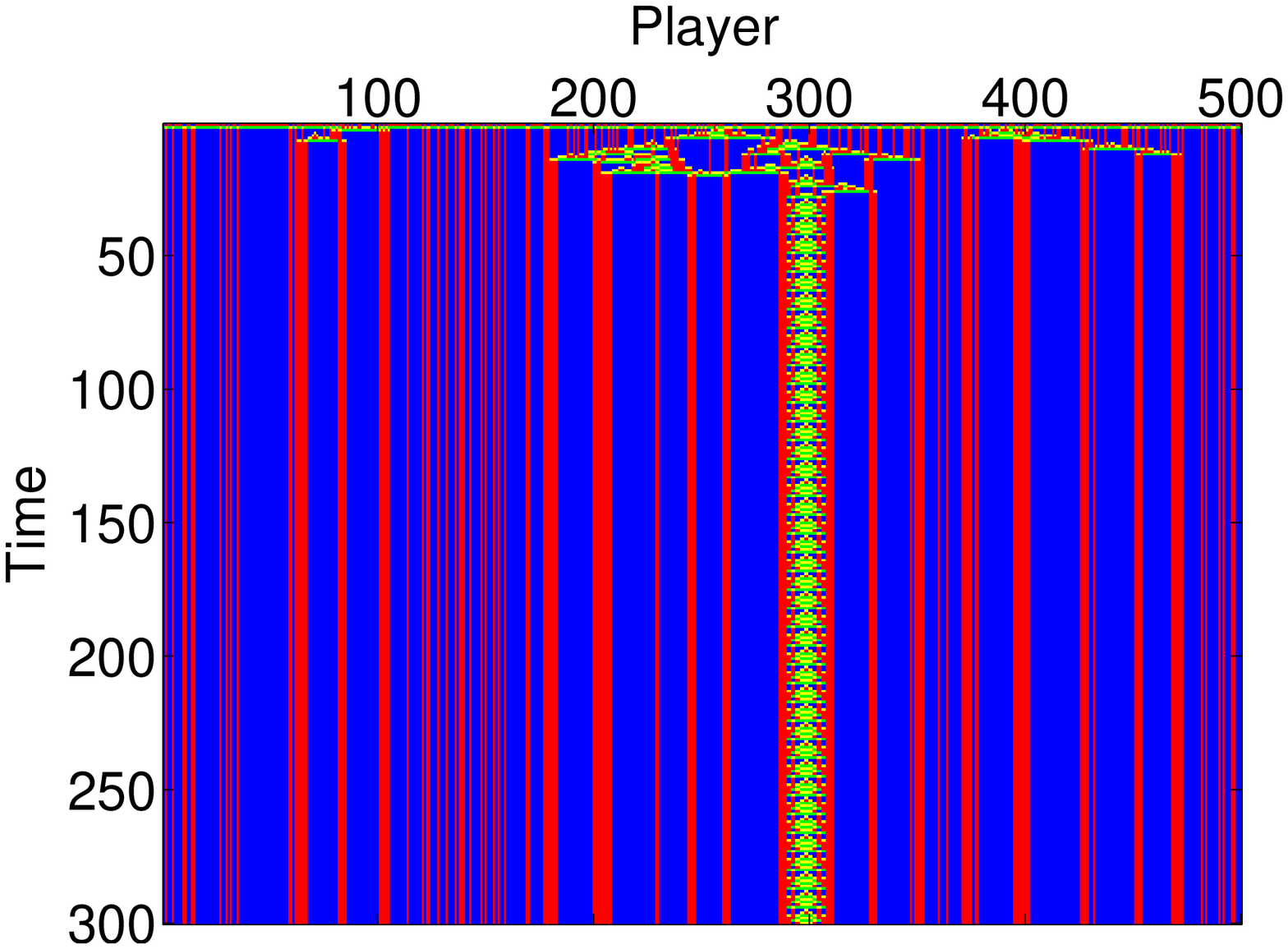}}
\subfloat[$T=1.40$]{\includegraphics[width=0.45\linewidth]{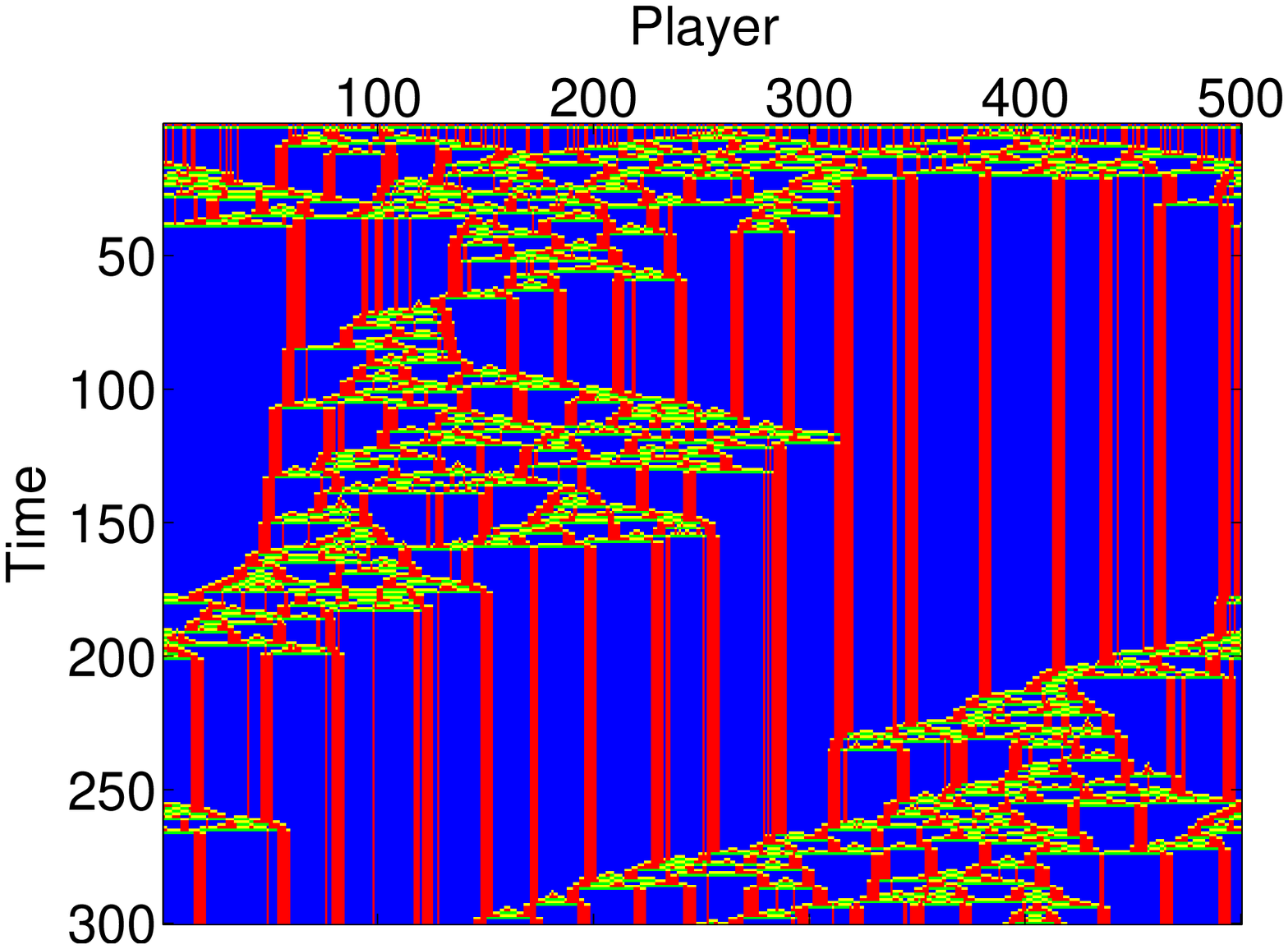}}
\\
\subfloat[$T=1.41$]{\includegraphics[width=0.45\linewidth]{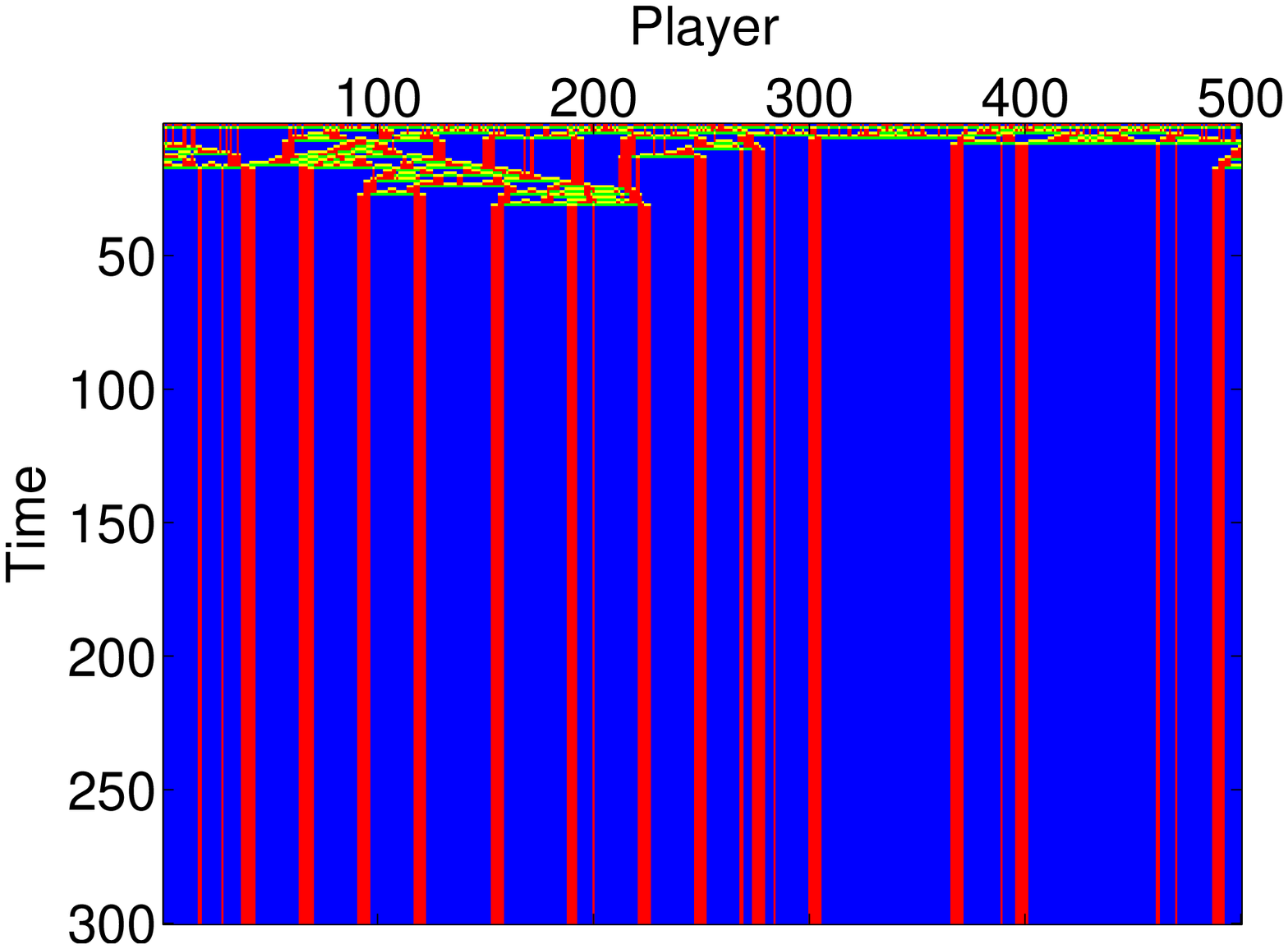}}
\subfloat[$T=1.67$]{\includegraphics[width=0.45\linewidth]{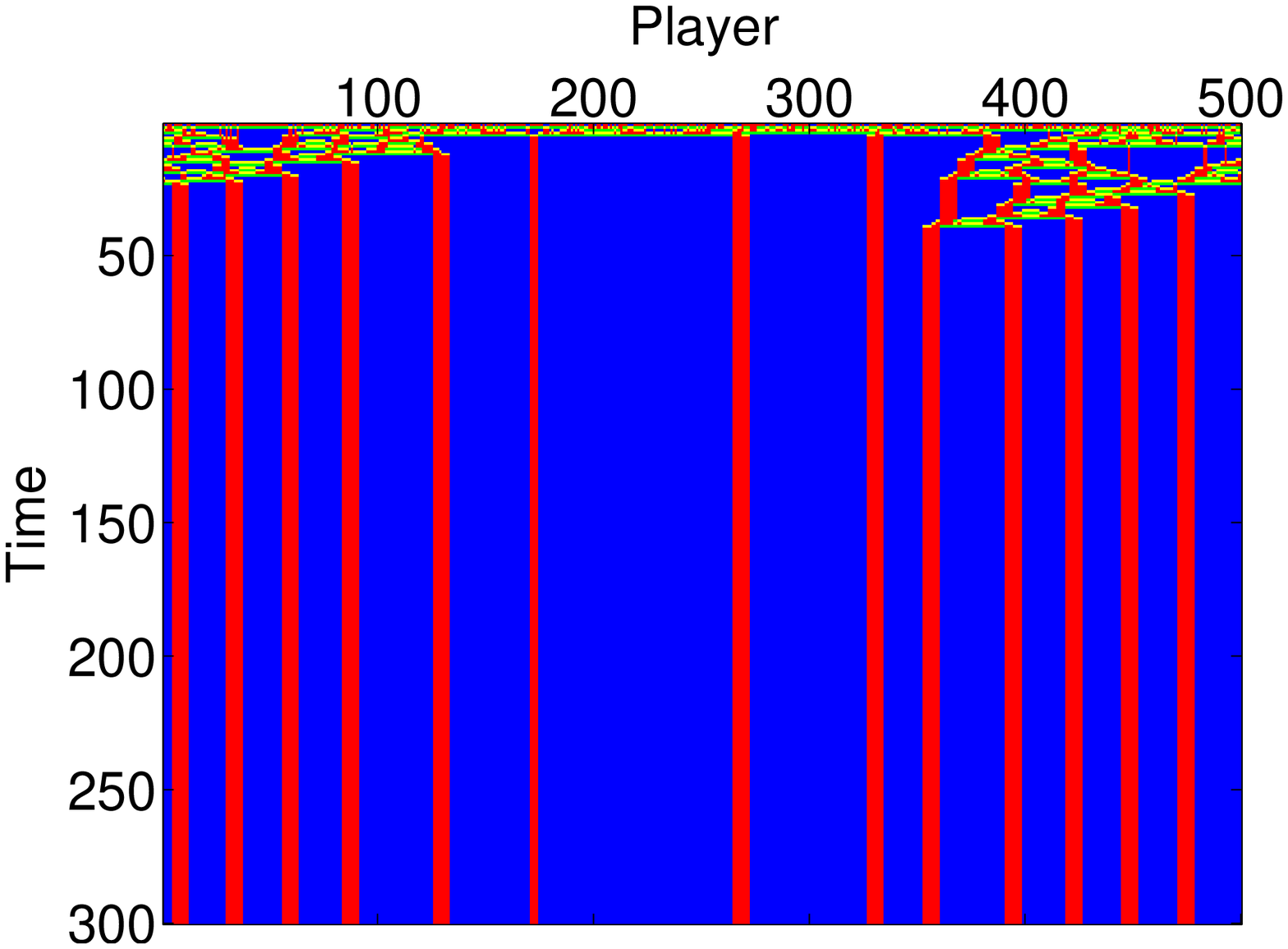}}
\\
\subfloat[$T=2.00$]{\includegraphics[width=0.45\linewidth]{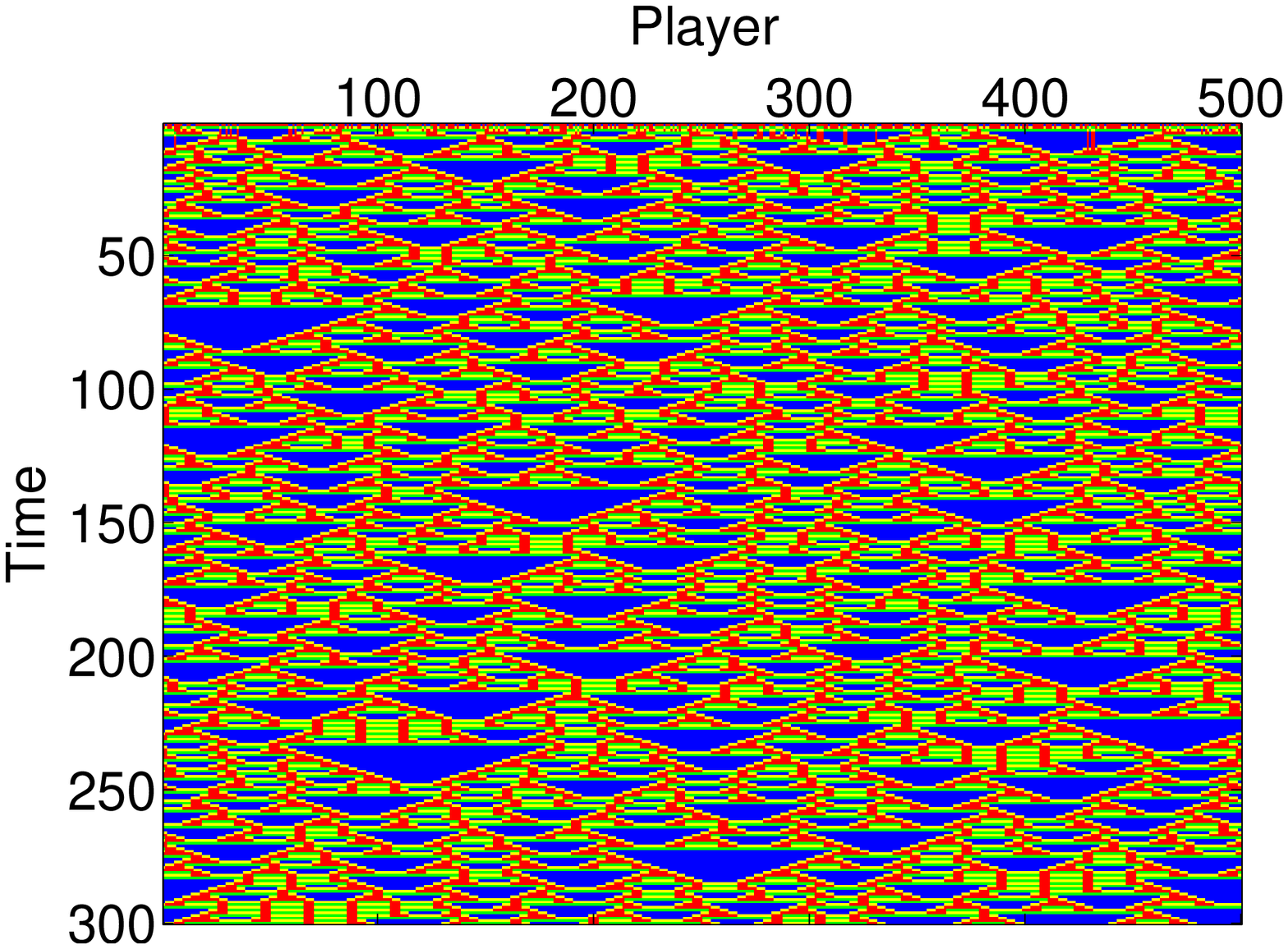}}
\caption{
Sequence of numerical simulations that show how the temptation $T$ change alter the cooperative/defective clusters patterns.
In the interval between the presented $T$ values do not occur changes in patterns.
The parameters of these simulations are: 
$L=500$, $t=300$, $\rho_0=0.3$ and $z=24$ (without self-interaction).
To: 
(a) $T=1.00$;
(b) $T=1.01$;
(c) $T=1.19$;
(d) $T=1.40$;
(e) $T=1.41$;
(f) $T=1.67$; and
(g) $T=2.00$.}
\label{fig_pav_serie}
\end{figure}

There are $T$ intervals where increases or decreases in its value do not alter the system dynamics and $\rho_\infty$ for a same system (identical cooperators initial configuration and $z$).
In Fig.~\ref{fig_pav_serie}, for $z = 24$, when the system passes through the critical temptation values $T_c=\{1;~13/11;~7/5;~15/11;~8/5;~17/9\}$, transient changes and patterns increase in quantity and variety, these are the phase transitions.
For instance, from $T=13/11$ to $T=1.19$ (see Figs. \ref{fig_pav_serie}c), appears some {\it gliders} in the initial steps and a complex {\it finger} emerges and propagate during all the system evolution.
From $T=1.39$ to $T=7/5$ (see Figs. \ref{fig_pav_serie}d) the initial {\it gliders} are increased and they propagate during time evolution, not emerging the {\it finger} as occured before.
Furthermore, for $1 \leq T < 2$, the system presents the cooperative phase in the steady regime and for $T = 2$ (see Fig. \ref{fig_pav_serie}g), the system enters in the {\it quasi}-regular phase.

For systems adopting the PES, analyzing the clusters patterns we see that the evolution depends strongly on the neighborhood composition of the cluster.
However, the player location in this neighborhood (configuration) is irrelevant for the total payoff determination.
We have noticed also that the transient and steady regimes depend on the system parameters, when $T$ varies, the transient regime duration changes.
In the steady regime, there are changes in the $\rho_\infty$ value when $T$ passes through the $T_c$ values.
A system can present the cooperative or {\it quasi}-regular phases.
In the {\it quasi}-regular phase, the system can yield a transient and after it achieves a periodic $\rho_\infty$ oscillation.

\section{Conclusion} \label{conclusao}
The one-dimensional cellular automata, where each cell is a player who plays the Prisoner Dilemma with $z$ neighbors, adopting the Pavlovian Evolutionary Strategy have been explored here.
We have obtained the analytical value of $T_c$.
Using numerical results we have validated the existence of the phases transition, that have been analytically calculated.
We also analyzed the stationary state of the system and explained the patterns of the clusters due the local interactions of players, given rise to a new phase, the {\it quasi}-regular one.

In short, our results are:
(i) phases transitions occur in well defined values of temptation $T_c$ that were defined analyticaly;
(ii) existence of the cooperative and {\it quasi}-regular phases, which depend on the temptation value to defect;
(iii) absence of the defective and chaotic phases;
(iv) the Tragedy of the Commons does not take place;
(v) the cooperation is more remarkable than in systems that adopt DES.

The mean payoff of players is greater when the players are concerned only with their own payoff.
If the players copy the action of the neighbor who received the largest payoff, they may worsen the outcome of the whole system.
Thus, the comparison and therefore, the greed by the greatest payoff, can cause the ruin of all.
In the situation where there is no way to coordinate the moves of players, the best would that everyone seek to have a positive payoff, even this positive payoff is not the maximum possible.
Thus they could maximize the payoff of the population as a whole.


\section*{Acknowledgments}
The authors have greatly profited from the discussions with H. Fort and R. da Silva.
M. A. P. would like to thank CAPES and CNPq for the fellowships.
A. S. M. acknowledges the agencies CNPq (303990/2007-4, 476862/2007-8) and PROSUL Project (490440/2007-0) for support.

\appendix
\section{Patterns formation} \label{app_pattern}
\begin{figure}[!htbp]
\centering
\includegraphics[width=1.0\linewidth]{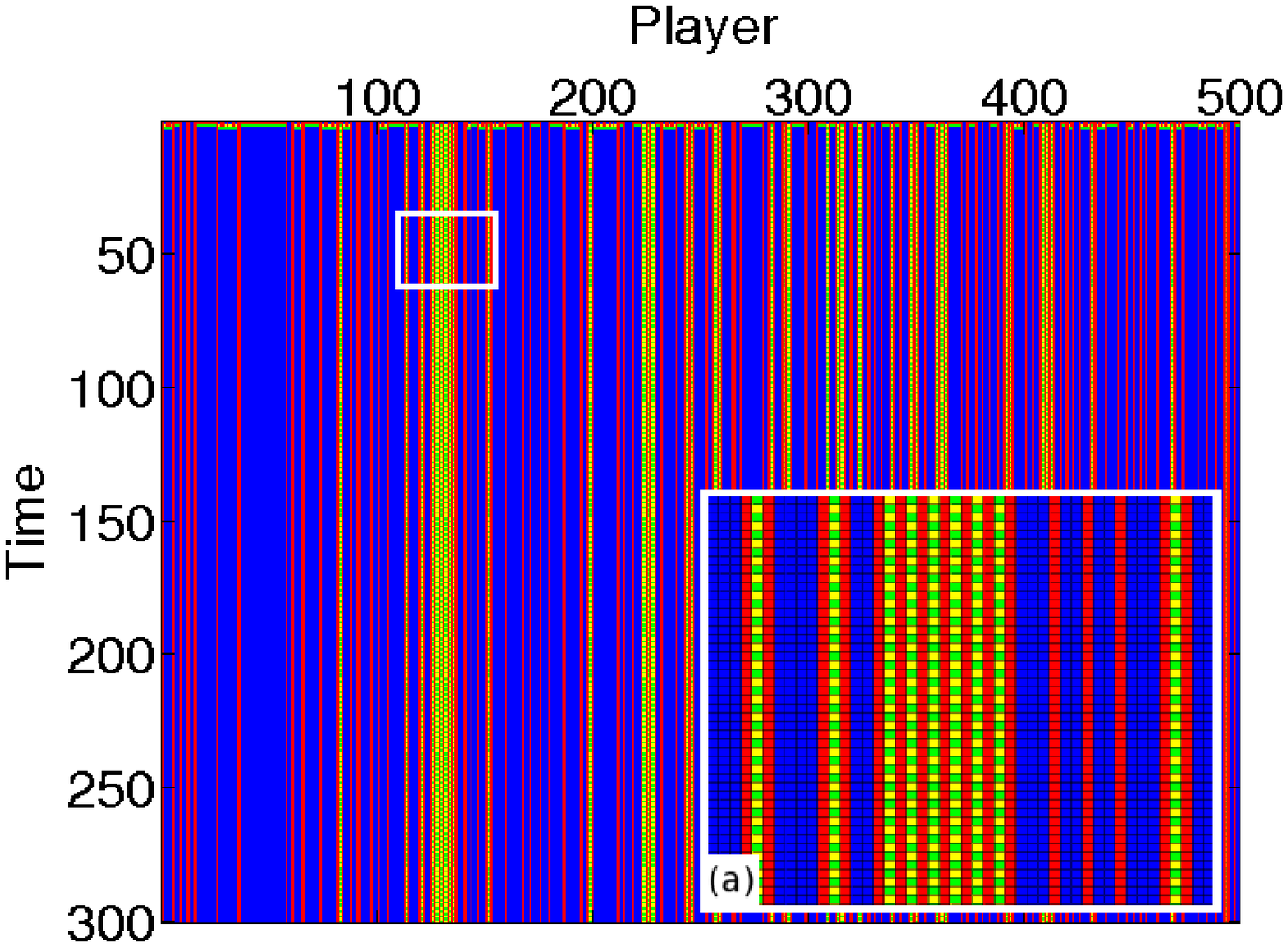}
\\
\vspace{0.8cm}
\includegraphics[width=0.7\linewidth]{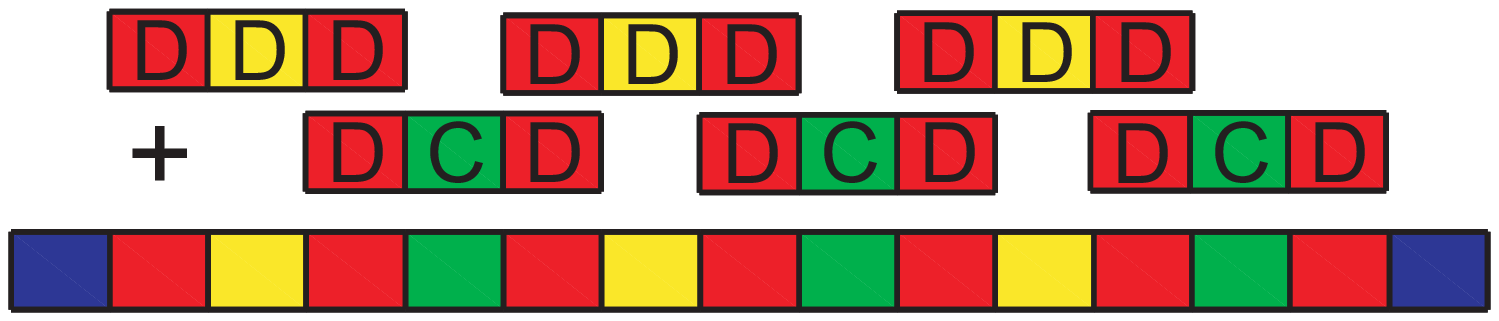}
\\ (b)
\caption{
Formation and evolution of smooth and complex {\it fingers}.
Parameters of this simulation are:
$L = 500$, $t = 300$, $T = 2.00$, $\rho_0 = 0.3$ and $z = 3$ (with self-interaction).
(a) zoom of the marked area.
(b) Formation of a pattern composed from elementary patterns.
}
\label{fig_pav_finger_1}
\end{figure}

In the following we analyze particular cases to explain the dynamics of evolution of these systems.
In the zoom of the images of Fig.~\ref{fig_pav_finger_1}, one sees that the complex {\it finger} with three players have the pattern\footnote{C: cooperator player, D: defector one. The player at center of the pattern is printed in boldface.}: \{D {\bf D} D $\rightarrow$ D {\bf C} D\} and with thirteen players the pattern is: \{D D D C D D D C D D D C D $\rightarrow$ D C D D D C D D D C D D D\}.
The pattern of 13 players is a composition formed by the alternation of the patterns of three players, with overlap (see Fig.~\ref{fig_pav_finger_1}b).
In other words, the patterns \{D {\bf D} D\} and \{D {\bf C} D\} combine themselves so that the third player of one pattern is the first of the following one.
Other combinations formed by the addition of patterns with or without overlapping of edges can be observed.

\begin{figure}[!htbp]
\centering{\includegraphics[width=1.0\linewidth]{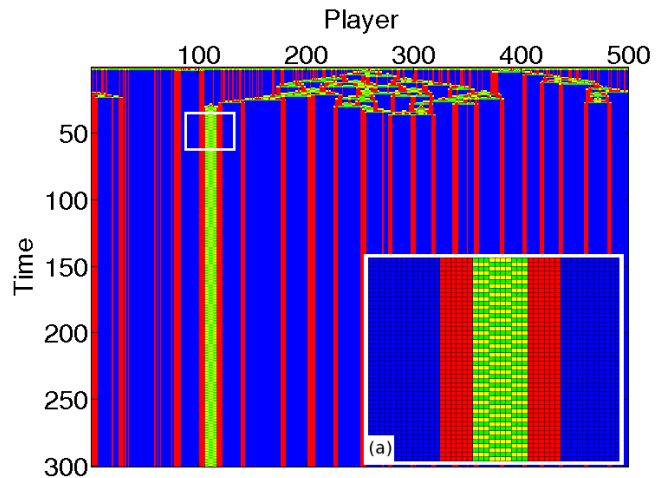}}
\caption{
Formation and evolution of smooth and complex fingers.
The parameters of this simulation are:
$L = 500$, $t = 300$, $T = 1.30$, $\rho_0 = 0.3$, e $z = 22$ (without self-interaction).
(a) zoom of the marked area.}
\label{fig_pav_finger_2}
\end{figure}

\begin{figure}[!htbp]
\centering
\includegraphics[width=1.0\linewidth]{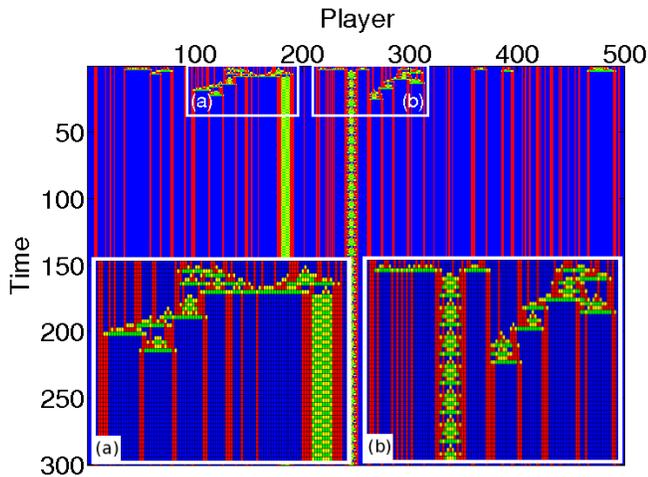}
\caption{
Formation and evolution of smooth and complex {\it fingers}.
The parameters of this simulation are:
$L = 500$, $t = 300$, $T = 1.10$, $\rho_0 = 0.7$, e $z = 12$ (without self-interaction).
(a) and (b) zoom of the marked areas (a) and (b), respectively.}
\label{fig_pav_finger_3}
\end{figure}

In the zoom of Fig.~\ref{fig_pav_finger_2} there are simple {\it fingers} with twelve defectors at most and also a complex one with the pattern: \{6D 3C {\bf 4D} 3C 6D $\rightarrow$ 6D 3D {\bf 4C} 3D 6D\}.
In the zooms of Fig.~\ref{fig_pav_finger_3}, emerging {\it fingers} have the pattern:
Fig.~\ref{fig_pav_finger_3}a: \{4D C 3D 3C D 4D $\rightarrow$ 4D D 3C 3D C 4D\} and
Fig.~\ref{fig_pav_finger_3}b: \{C D 2C {\bf D} 2C D C $\rightarrow$ C D C {\bf 3D} C D C $\rightarrow$ 3D {\bf 3C} 3D $\rightarrow$ 2D C {\bf 3D} C 2D $\rightarrow$ D C D {\bf 3C} D C D $\rightarrow$ D C {\bf 5D} C D $\rightarrow$ C D {\bf 5C} D C\}.
Note the periodicity present in these patterns.

\bibliographystyle{unsrt}
\bibliography{bibliografia_pavlov}

\begin{thebibliography}{10}

\bibitem{axelrod_1984}
R.~Axelrod.
\newblock {\em The evolution of cooperation}.
\newblock Basic Books, New York, 1984.

\bibitem{axelrod_1981}
R.~Axelrod and W.~D. Hamilton.
\newblock The evolution of cooperation.
\newblock {\em Science}, 211:1390--1396, 1981.

\bibitem{anteneodo_2002}
C.~Anteneodo, C.~Tsallis, and A.~S. Martinez.
\newblock Risk aversion in economic transactions.
\newblock {\em Europhys. Lett.}, 59(5):635--641, 2002.

\bibitem{stauffer_2004}
D.~Stauffer.
\newblock Introduction to statistical physics outside physics.
\newblock {\em Physica A}, 336:1--5, 2004.

\bibitem{bouchaud_2002}
J.~P. Bouchaud.
\newblock An introduction to statistical finance.
\newblock {\em Physica A}, 313:238--251, 2002.

\bibitem{turner_1999}
P.~E. Turner and L.~Chao.
\newblock Prisoner´s dilemma in an rna virus.
\newblock {\em Nature}, 398:441--443, 1999.

\bibitem{pincus_1970}
M.~Pincus.
\newblock An evolutionary strategy.
\newblock {\em J. theor. Biol.}, 28:483--488, February 1970.

\bibitem{nowak_1992}
M.~A. Nowak and R.~M. May.
\newblock Evolutionary games and spatial chaos.
\newblock {\em Nature}, 359:826--829, 1992.

\bibitem{fort_2005}
H.~Fort and S.~Viola.
\newblock Spatial patterns and scale freedom in prisoner's dilemma cellular
  automata with pavlovian strategies.
\newblock {\em J. Stat. Mech.-Theory Exp.}, 1(P01010), 2005.

\bibitem{beyer_2002}
H.~G. Beyer and H.~P. Schwefel.
\newblock Evolution strategies.
\newblock {\em Natural Computing}, 1:3--52, 2002.

\bibitem{thorndike_1911}
Edward~L. Thorndike.
\newblock {\em Animal Intelligence}.
\newblock Macmillan, 1911.

\bibitem{kraines_1989}
D.~Kraines and V.~Kraines.
\newblock Pavlov and the prisoner's dilemma.
\newblock {\em Theory and Decision}, 26:47--79, 1989.

\bibitem{posch_1999}
M.~Posch.
\newblock Win-stay, lose-shift strategies for repeated games - memory lenght,
  aspiration levels and noise.
\newblock {\em J. theor. Biol.}, 198:183--195, 1999.

\bibitem{kraines_1993}
D.~Kraines and V.~Kraines.
\newblock Learning to cooperate with pavlov - an adaptive strategy for the
  iterated prisoners-dilemma with noise.
\newblock {\em Theory Decis.}, 35(2):107--150, 1993.

\bibitem{kraines_1995}
D.~Kraines and V.~Kraines.
\newblock Evolution of learning among pavlov strategies in a competitive
  environment with noise.
\newblock {\em J. Confl. Resolut.}, 39(3):439--466, 1993.

\bibitem{lorberbaum_2002}
J.~P. Lorberbaum, D.~E. Bohning, A.~Shastri, and L.~E. Sine.
\newblock Are there really no evolutionarily stable strategies in the iterated
  prisoner's dilemma?
\newblock {\em J. Theor. Biol.}, 214(2):155--169, 2002.

\bibitem{nowak_1993}
M.~Nowak and K.~Sigmund.
\newblock A strategy of win stay, lose shift that outperforms tit-for-tat in
  the prisoners-dilemma game.
\newblock {\em Nature}, 364(6432):56--58, Jul 1993.

\bibitem{soares_2006}
R.~O.~S. Soares and A.~S. Martinez.
\newblock The geometrical pattern of the evolution of cooperation in the
  spatial prisoner's dilemma: an intra-group model.
\newblock {\em Physica A}, 369:823--829, 2006.

\bibitem{pereira_2008_IJMPC}
M.~A. Pereira, A.~S. Martinez, and A.~L. Esp\'{i}ndola.
\newblock Prisoner's dilemma in one-dimensional cellular automata:
  Visualization of evolutionary patterns.
\newblock {\em Int. J. of Modern Phys. C}, 2008.

\bibitem{pereira_2008_BJP}
M.~A. Pereira, A.~S. Martinez, and A.~L. Esp\'{i}ndola.
\newblock An exhaustive exploration of the parameter space of the prisoners'
  dilemma in one-dimensional cellular automata.
\newblock {\em Brazilian Journal Of Physics}, 38(1):65--69, March 2008.

\bibitem{duran_2005}
O.~Dur\'{a}n and R.~Mulet.
\newblock Evolutionary prisoner's dilemma in random graphs.
\newblock {\em Physica D}, 208:257--265, 2005.

\bibitem{dresher_1961}
M.~Dresher.
\newblock {\em The Mathematics of Games of Strategy: Theory and Applications}.
\newblock Prentice-Hall, Englewood Cliffs, NJ, 1961.

\bibitem{ifti_2004}
M.~Ifti, T.~Killingback, and M.~Doebeli.
\newblock Effects of neighbourhood size and connectivity on the spatial
  continuous prisoner's dilemma.
\newblock {\em J. Theor. Biol.}, 231:97--106, 2004.

\bibitem{nowak_1999b}
Lindi~M. Wahl and Martin~A. Nowak.
\newblock The continuous prisoner's dilemma: I. linear reactive strategies.
\newblock {\em J. theor. Biol.}, 200:307--321, 1999.

\bibitem{nowak_1999c}
Lindi~M. Wahl and Martin~A. Nowak.
\newblock The continuous prisoner's dilemma: Ii. linear reactive strategies
  with noise.
\newblock {\em J. theor. Biol.}, 200:323--338, 1999.

\end{thebibliography}

\end{document}